\documentclass[12pt,3p,authoryear]{elsarticle}
\usepackage[T1]{fontenc}
\usepackage{amsmath,amssymb,amsfonts}
\usepackage{natbib}
\usepackage{graphicx}
\usepackage{epstopdf}
\usepackage[hidelinks,breaklinks=true]{hyperref}
\usepackage{booktabs,geometry}
\usepackage{rotating}
\usepackage{times}
\usepackage{xspace}
\usepackage{dcolumn}
\newcolumntype{.}[1]{D{.}{.}{#1}}
\newcolumntype{,}[1]{D{,}{,}{#1}}
\usepackage{setspace}
\usepackage{color}
\usepackage{subdepth}
\usepackage{needspace}   
\usepackage{float}
\usepackage[font={small,singlespacing},labelfont={bf,small},skip=0cm]{caption}
\floatstyle{plaintop}
\restylefloat{table}
\restylefloat{figure}			
\usepackage{rotating}
\usepackage{multirow,adjustbox}
\DeclareMathAlphabet\mathbfcal{OMS}{cmsy}{b}{n}
\usepackage[font={normal}]{subfig}
\usepackage{dsfont}

\usepackage[table]{xcolor}
\newcommand{\grbb}[1]{\textbf{#1}}

\usepackage{amsmath, amssymb, amscd, amsthm, etex, etoolbox, euscript, enumerate,
graphics, hhline, dsfont, multicol, multirow, tabularx}	

\journal{arXiv}

\newcommand{\ie}{\emph{i.e.}\xspace}
\newcommand{\eg}{\emph{e.g.}\xspace}

\newtheorem{theorem}{Theorem}
\newtheorem{proposition}{Proposition}
\newtheorem{assumption}{Assumption}
\newtheorem{remark}{Remark}
\newtheorem{lemma}{Lemma}

\usepackage{color}

\def\bb{\mathbf b}

\def\bg{\mathbf g}

\def\bi{\mathbf i}

\def\bm{\mathbf m}

\def\bp{\mathbf p}

\def\br{\mathbf r}

\def\bv{\mathbf v}
\def\bw{\mathbf w}
\def\bx{\mathbf x}

\def\bB{\mathbf B}

\def\bD{\mathbf D}
\def\bE{\mathbf E}

\def\bG{\mathbf G}

\def\bI{\mathbf I}

\def\bO{\mathbf 0}

\def\bQ{\mathbf Q}
\def\bR{\mathbf R}

\def\bV{\mathbf V}

\def\bX{\mathbf X}
\def\bY{\mathbf Y}

\def\balpha{\boldsymbol{\alpha}}
\def\bbeta{\boldsymbol{\beta}}
\def\bgamma{\boldsymbol{\gamma}}
\def\bdelta{\boldsymbol{\delta}}

\def\etab{\boldsymbol{\eta}}
\def\btheta{\boldsymbol{\theta}}

\def\blambda{\boldsymbol{\lambda}}
\def\bmu{\boldsymbol{\mu}}
\def\bnu{\boldsymbol{\nu}}

\def\bomega{\boldsymbol{\omega}}
\def\bGamma{\boldsymbol{\Gamma}}

\def\bTheta{\boldsymbol{\Theta}}

\def\bSigma{\boldsymbol{\Sigma}}

\begin{document}
\begin{frontmatter}
\title{High-Dimensional Mean-Variance Spanning Tests\\
\vspace{1cm}
\Large{\emph{Preliminary}}}

\author[hec]{David Ardia}
\ead{david.ardia@hec.ca}
\author[aix]{Sébastien Laurent}
\ead{sebastien.laurent@univ-amu.fr}
\author[hec]{Rosnel Sessinou\corref{cor1}}
\cortext[cor1]{Corresponding author. HEC Montréal, 3000 Chemin de la Côte-Sainte-Catherine, Montreal, QC H3T 2A7.}
\ead{rosnel.sessinou@hec.ca}
\address[hec]{GERAD \& Department of Decision Sciences, HEC Montréal, Montréal, Canada}
\address[aix]{Aix-Marseille School of Economics, CNRS \& EHESS, Aix-Marseille Graduate School of Management – IAE, France}

\begin{abstract}
We introduce a new framework for the mean-variance spanning (MVS) hypothesis testing. The procedure can be applied to any test-asset dimension and only requires stationary asset returns and the number of benchmark assets to be smaller than the number of time periods. It involves individually testing moment conditions using a robust Student-t statistic based on the batch-mean method and combining the
p-values using the Cauchy combination test. Simulations demonstrate the superior performance of the test compared to state-of-the-art approaches. For the empirical application, we look at the problem of domestic versus international diversification in equities. We find that the advantages of diversification are influenced by economic conditions and exhibit cross-country variation. We also highlight that the rejection of the MVS hypothesis originates from the potential to reduce variance within the domestic global minimum-variance portfolio.
\end{abstract}
\end{frontmatter}

\doublespacing
\newpage

\newpage
\section{Introduction}

Mean-variance spanning (MVS) tests aim to determine if adding test assets to a set of benchmark assets improves the mean-variance efficient frontier. The two-fund separation theorem suggests that MVS testing is equivalent to testing if test assets have zero weights in the maximum Sharpe ratio and global minimum-variance portfolios of all assets. These tests are called ``spanning tests,'' and rejecting any of them leads to the rejection of the MVS hypothesis. These tests can be performed individually. When the focus is on the maximum Sharpe ratio portfolio of the asset excess returns only, and the market factors are the benchmark assets, the term ``mean-variance efficiency'' test is used instead of the spanning test. Such spanning tests are relevant for validating linear factor asset pricing models, such as the CAPM \citep{Sharpe1964,Lintner1965} or the three-factor model by \citet{FamaFrench1993}. For simplicity, we use the generic term MVS to refer to all these tests.

Several MVS tests exist in the literature, but they all have limitations; see \citet{DeRoonNijman2001} and \citet[][Chapter 16]{FrancoisHubner2024} for a review. Some of them \citep[\eg,][]{huberman1987mean, gibbons1989test, britten1999sampling, kempf2006estimating} require the estimation of a precision matrix, which means that the number of test assets must be much smaller than the number of time periods $T$. This problem can be avoided by testing with ad-hoc portfolios instead of individual assets, but this introduces an aggregation bias \citep[\eg,][]{roll1977critique} and does not ensure asset-level results. Some others \citep[\eg,][]{beaulieu2007multivariate, beaulieu2010asset, GungorLuger2009} use simulations to relax the normality assumption in previous studies and deal with serial dependence. \cite{kan2012tests} also introduced a score test to cope with the same problem. However, these approaches require a small test set to be used. \citet{PesaranYamagata2012} then developed an MVS test that can handle large test sets, but only under weak and sparse correlation in the disturbances. \citet{gungor2016multivariate} show that their test loses power as the correlations between the assets increase. We also find that that test is oversized in the realistic setting where asset returns follow AR-GARCH processes. \citet{gungor2016multivariate} finally introduced a simulation-based method for MVS with a large test set. However, we find their test becomes uninformative when the number of benchmark assets is large despite being lower than the number of time periods. 

In this paper, we propose a new framework for MVS testing that can be applied to any test-asset dimension and only requires stationary asset returns. We use new moment conditions for spanning and test them in two steps. First, we test each component of the moment vector using robust Student-t tests based on the batch-mean method. We improve the computational efficiency of the method by avoiding refitting the same model many times. Second, we combine the individual p-values using the CCT of \citet{liu2020cauchy}, which accounts for the cross-sectional dependence between the test statistics and is valid under weak assumptions. We prove that the CCT can combine Student-t test statistics p-values, and we show that it has good finite sample properties. Monte Carlo simulations shows that our MVS tests have correct size and high power in most setups. They also work well on heteroscedastic, skewed, and fat-tailed data. However, we find that when we rely on a standard batch-mean method, our test suffers from size distortion when the number of benchmark assets is large. We solve this problem by using a randomly-weighted batch-mean procedure on our new moment conditions. 

For the empirical illustration, we apply our test to determine whether combining blue-chip stocks traded in the U.S., Canada, and Europe can improve each country's domestic mean-variance efficient frontier. Previous studies used portfolio-based MVS tests to deal with the high-dimensional nature of the problem, but results are subject to aggregation bias \citep{huberman1987mean,britten1999sampling,bekaert1995time,kempf2006estimating}. We use our test at the asset level and find that the benefits of international diversification depend on economic conditions and vary across countries, in line with the literature. The MVS hypothesis is rejected by the variance reduction potential in the domestic global minimum-variance portfolio.

%

The paper is structured as follows. Section \ref{sec:metho} introduces our new testing procedure. Section \ref{sec:mcsim}  presents an overview of MVS tests and studies the finite sample performance of these tests and our newly proposed tests using Monte Carlo simulations. Section \ref{sec:empirics} presents the empirical application, and Section \ref{sec:conclusion} concludes.

\section{Testing MVS Assumption With Many Test Assets and Serial Dependence}\label{sec:metho}

\subsection{The Limits of the Existing Framework}

Define $\br_t \equiv ( \br_{1,t}', \br_{2,t}' )' \equiv (r_{1,t},\ldots,r_{K,t},r_{K+1,t},\ldots,r_{K+N,t})'$ as the vector of returns of the $K+N$ risky assets at time $t$, where $\br_{1,t}$ is the vector of returns of the $K$ benchmark assets and $\br_{2,t}$ is the vector of returns of the $N$ test assets. Define the expected value and covariance matrix of $\br_t$ as $\bmu \equiv \mathbb E[\br_t] $ and $\bV\equiv\mathbb{V}[\br_{t}]$, respectively. We assume $\bV$ to be nonsingular. Let $\bQ \equiv [\bO_{N \times K} \, \vdots \, \bI_{N}]$ be a selection matrix where $\bO_{N \times K}$ is a $N \times K$ matrix of zeros, $\bI_{N}$ an identity matrix of size $N$, and $\bi_{N}$ a column vector of ones of length $N$. 

Several MVS tests have been proposed in the literature and we consider, below, the three most common hypotheses. 
\begin{itemize}
\item By the two-fund separation theorem, the global null hypothesis of MVS can be expressed as
\begin{equation}\label{H0MVS_both}
H_0^{\balpha,\bdelta}: \bQ [{\bw^{\balpha}} \, \vdots \, {\bw^{\bdelta}} ]  = \bO_{N \times 2} \,,
\end{equation}
where  $\bw^{\balpha}\equiv\frac{\bV^{-1} \bmu}{\bi'_{N+K} \bV^{-1}\bmu} $ and  $\bw^{\bdelta}\equiv\frac{\bV^{-1} \bi_{N+K}}{\bi'_{N+K} \bV^{-1} \bi_{N+K}}$ denote respectively the vectors of weights of the maximum Sharpe and the global minimum-variance portfolios. 
\item The null hypothesis of maximum Sharpe portfolio spanning can be expressed as 
\begin{equation}\label{H0MVS_alpha}
H_0^{\balpha}: \bQ\bw^{\balpha}=\bO_{N \times 1}\,.
\end{equation}
\item The global minimum-variance portfolio spanning hypothesis is 
\begin{equation}\label{H0MVS_delta}
H_0^{\bdelta}: \bQ\bw^{\bdelta}=\bO_{N \times 1} \,. 
\end{equation}
\end{itemize}

The test of $H_0^{\balpha}$ is referred to as a mean-efficiency test in the literature when excess returns (over the risk-free rate) of the test assets are used and the benchmark assets are factors (such as the market, size, and value factors in the \citet{FamaFrench1993} model). Importantly, \citet{huberman1987mean}, \citet{britten1999sampling}, and \citet{kempf2006estimating} show that testing the above three spanning hypotheses can be done using tests of linear restrictions on the parameters of linear regression models.

Table~\ref{tab:hypo} shows, for the three hypotheses, the corresponding restrictions on the weights, the linear models used to test these hypotheses, the assumptions made on the linear model, the linear restrictions corresponding to these hypotheses, and the key references linking the restrictions on the weights and the linear restrictions. 

\begin{table}[H]
\caption{\textbf{Testing Spanning Hypotheses via Tests of Linear Restrictions on Linear Regression Models}\\
For any $d$-dimensional vector $\bv\equiv(v_1,v_2,\ldots,v_d)'$, $\bv_{-1}\equiv(v_2,\ldots,v_d)'$. Recall that $\bi_{N}'\bw^{\alpha}=\bi_{N}'\bw^{\delta}=1$. We let  $\bQ \equiv [\bO_{N \times K} \, \vdots \, \bI_{N}]$ and $\bQ_{-1} \equiv [\bO_{N \times {(K-1)}} \, \vdots \, \bI_{N}]$.  Note that $\bQ\bw^{\delta'}=\bQ_{-1}\bw^{\delta'}_{-1}$  is a column vector with elements $w^{\delta'}_{j}$ for $j=K+1,\ldots,N+K$.}
\label{tab:hypo}
\singlespacing
\adjustbox{width=\textwidth}{
\begin{tabular}{l|c|c|c}
\toprule
Hypothesis & $H_0^{\balpha,\bdelta}$ & $H_0^{\balpha}$ & $H_0^{\bdelta}$ \\
\midrule
Restrictions on weights & $\bQ [{\bw^{\balpha}} \, \vdots \, {\bw^{\bdelta}} ]  = \bO_{N \times 2}$ & $\bQ\bw^{\balpha}=\bO_{N\times 1}$ & $\bQ\bw^{\bdelta}=\bO_{N\times 1}$ \\
Linear model & $\br_{2,t} = \balpha + \bbeta \br_{1,t} + \etab_t$ & $1=\bb_1' \br_{1,t}+\bb_2'\br_{2,t}+u_t$ & $r_{1,1,t}=c+\bw^{\delta'}_{-1} (\bi_{N+K-1}r_{1,1,t}-\br_{-1,t})+e_t$ \\
Assumptions & $\mathbb E[\etab_t]=\bO_{N\times 1}$, $\mathbb E[\br_{1,t}\etab_t']={\bO_{K \times N}}$ & $\mathbb{E}((1,\br_t')'u_t)=\bO_{(K+N) \times 1}$ & $\mathbb{E}[(1,(\bi_{N+K-1}r_{1,1,t}-\br_{-1,t})')'e_t]=0$ \\
Linear restrictions & $\balpha=\bO_{N \times 1}; \bdelta\equiv\bi_N-\bbeta \bi_K=\bO_{N \times 1}$ & $\bb_2=\bO_{N \times 1}$ & $\bQ_{-1}\bw_{-1}^\delta=\bO_{N \times 1}$ \\
References & \citet{huberman1987mean} & \citet{britten1999sampling} & \citet{kempf2006estimating} \\
\bottomrule
\end{tabular}}
\end{table}

Parameter $\bbeta$ is a $N\times K$ matrix with elements $(\beta_{i,j})$ while $\bbeta_{\bullet,-1}$ denotes the $N\times (K-1)$ submatrix of $\bbeta$ whose first column has been removed and let $\bdelta\equiv\bi_N-\bbeta \bi_K$. The model $\br_{2,t} = \balpha + \bbeta \br_{1,t} + \etab_t$
can be rewritten as 
\begin{equation}
\br_{2,t}- \bi_{N}r_{1,1,t} = \balpha + \bdelta r_{1,1,t}+ \bbeta_{\bullet,-1} (\br_{1,-1,t}- \bi_{K-1}r_{1,1,t}) + \etab_t
\label{eq_tests}
\end{equation}
so that $\bdelta$ appears in the model. Testing $H_0^{\balpha}$,  $H_0^{\bdelta}$ and $H_0^{\balpha,\bdelta}$ can therefore be done by testing respectively
$H_0: \balpha=\bO_N$, $H_0: \bdelta=\bO_N$  and $H_0: \balpha=\bO_N; \bdelta=\bO_N$ on Model~\eqref{eq_tests}.

The most natural way to test these hypotheses would be to assume that $\etab_t$ is Gaussian and to use F-tests as advocated by \citet{huberman1987mean}, \citet{britten1999sampling}, and \citet{kempf2006estimating}. F-tests for the above hypotheses have been shown to have good finite sample properties when the number of test assets $N$ and the number of benchmark assets $K$ are much lower than the number of time periods $T$ and when residuals are \textit{i.i.d}. However, attention must be paid to the presence of serial correlation and heteroscedasticity in the residuals. 
\citet{lazarus2018har,lazarus2019size} and \citet{pedersen2020robust} have reported that standard HAC procedures can be size distorted in finite samples. Bootstrapping is an alternative, but it is computationally very demanding, especially when $N$ is large.

To keep $N$ small, spanning tests are generally applied at the portfolio level, not at the asset level. We will show that this strongly impacts the conclusions and recommend carrying out the test at the asset level. In this case, $N$ will most likely be large (even larger than $T$), which invalidates standard methods such as F-tests and score tests (see \citealp{kan2012tests} and subsequent references for details about those tests).

Our contribution is twofold. 

First, we propose to rely on the batch-mean method instead of HAC estimates of standard errors to robustify the statistical inference to the possible presence of serial correlation and heteroscedasticity in the residuals. According to \citet{pedersen2020robust}, the batch-mean method enjoys better finite sample properties than standard HAC procedures in the presence of serial correlation. This method requires estimating the model over sub-samples and performing a t-test on the collection of estimated coefficients. It also imposes the restriction that $K$ must be smaller than the size of each sub-sample, which must be of order $T^{2/3}$ according to \citet{sherman1997subseries} and \citet{flegal2010batch}. This prohibits using standard batch-mean when $K$ is of that order.
We also show how to overcome this restriction by reformulating the null hypotheses of spanning as residual-based moment conditions. This allows one to avoid refitting the model when using the batch-mean method. Furthermore, a weighted batch-mean method is presented to cope with the size distortion appearing for the standard batch-mean when $K$ is large but smaller than $T$.

Second, we propose to apply the tests to each test asset independently, that is, by testing for $j=1,\ldots,N$ the null hypotheses $H^{\alpha_j}_0: \alpha_j=0$, $H^{\delta_j}_0: \delta_j=0$  and/or $H^{\alpha_j,\delta_j}_0: \alpha_j=0; \delta_j=0$  
and collecting the corresponding p-values. Then, to test the join hypotheses on the $N$ assets, we propose to aggregate transformed individual p-values into a global p-value using the Cauchy Combination Test (CCT) of \citet{liu2020cauchy}. This avoids having to estimate the variance-covariance matrix of the $N$ estimates of $\alpha_j$ and/or $\delta_j$, which would be the case for a classic F-test and which would prove inefficient when $N$ is large. Unlike traditional p-value merging procedures such as the Bonferroni method or \citet{benjamini1995controlling} and \citet{benjamini2001control}'s methods, the CCT features the nice property that when all hypotheses are true, the empirical size converges to the nominal size as the significance level tends to zero (\eg, for a nominal size of 5\%), for arbitrary dependency structures among the test statistics.   	

\subsection{A New Framework for MVS Test}\label{illustration}

Proposition~\ref{th:recast} provides a rewriting of the null hypotheses $H_0^{\balpha}$ and $H_0^{\bdelta}$ using moment conditions on the residuals of regressions models.  An equivalent rewriting of $H_0^{\balpha,\bdelta}$ is deduced as a by-product.

\begin{proposition}~\label{th:recast} Let us define $\varepsilon(y_t,\bx_t,\bgamma)\equiv y_t-\bgamma \bx_t$, where $y_t\in\mathbb{R}$,  $\bx_t\in\mathbb{R}^q$  and $\bgamma\in\mathbb{R}^q$ such that $\mathbb{E}\left[\bx_t(y_t-\bgamma' \bx_t)\right]=\bO_{q \times 1}$. Let $\bx_{t}^j \equiv(x_{1,t}^j,\ldots, x_{K+2,t}^j) \equiv (r_{2,j,t}-r_{1,1,t},1,r_{1,1,t},\br_{1,-1,t}'-\bi_{K-1}' r_{1,1,t})'$, and denote $\bx_{-i,t}^j$ the vector $\bx_{t}^j$ from which $x_{i,t}^j$ is removed.	For $j=1,\ldots, N$, one has

\begin{enumerate}
\item[(a)] $ \mathbb{E}[g_{\alpha_j} (\br_{t},\btheta_j)] \equiv \mathbb{E}\left[\varepsilon(x_{1,t}^j,\bx_{-1,t}^j \,, \btheta_{j,1})\varepsilon(x_{2,t}^j,\bx_{-2,t}^j,\btheta_{j,2})\right]\propto\alpha_j$\,,
\item[(b)] $  \mathbb{E}[g_{\delta_j} (\br_{t},\btheta_j)] \equiv \mathbb{E}\left[\varepsilon(x_{1,t}^j,\bx_{-1,t}^j,\btheta_{j,1})\varepsilon(x_{3,t}^j,\bx_{-3,t}^j,\btheta_{j,3})\right]\propto\delta_j$ \,,
\end{enumerate}
with $\btheta_{j,i}\in\mathbb{R}^{K+1}$ for $i=1,2,3$ and $\btheta_j \equiv (\btheta_{j,1}',\btheta_{j,2}',\btheta_{j,3}')'\in\mathbb{R}^{3(K+1)}$.
\end{proposition}

Let us also define $\btheta \equiv  (\btheta_{1}',\ldots,\btheta_{N}')'\in\mathbb{R}^{3N(K+1)}$. Proposition \ref{th:recast} suggest that the null hypothesis of MVS can be written using 
moment conditions, that is:

\begin{itemize}
\item $H_{0}^{\balpha}:\; \balpha=\bO_{N\times 1}\Leftrightarrow \bm^{\balpha}(\btheta)=\mathbb{E}[\bg_{\balpha} (\br_{t},\btheta)]\equiv\mathbb{E}\left[\left(g_{\alpha_1} (\br_{t},\btheta_1),\ldots,g_{\alpha_N} (\br_{t},\btheta_N)\right)'\right]=\bO_{N\times 1}$,
\item $H_{0}^{\bdelta}:\; \bdelta=\bO_{N\times 1} \Leftrightarrow \bm^{\bdelta}(\btheta)=\mathbb{E}[\bg_{\bdelta} (\br_{t},\btheta)]\equiv\mathbb{E}\left[\left(g_{\delta_1} (\br_{t},\btheta_1),\ldots,g_{\delta_N} (\br_{t},\btheta_N)\right)'\right]=\bO_{N\times 1}$,
\item $H_{0}^{\balpha,\bdelta}:\; \balpha=\bdelta=\bO_{N\times 1}\Leftrightarrow \bm^{\balpha,\bdelta}(\btheta)=\mathbb{E}[\bg_{\balpha,\bdelta} (\br_{t},\btheta)]\equiv\mathbb{E}\left[\left(\bg_{\balpha} (\br_{t},\btheta)',\bg_{\bdelta} (\br_{t},\btheta)'\right)'\right]=\bO_{2N\times 1}.$
\end{itemize}

The proof of Proposition \ref{th:recast} is given in the online appendix. But for the sake of illustration, consider a universe of $K=2$ benchmark assets and $N=1$ test asset. As shown in column 2 of Table~\ref{tab:hypo}, when $N=1$, testing $H_0^{\alpha_1,\delta_1}$ implies testing 
$H_0: \alpha_1=0, \delta_1 \equiv 1-\beta_{1,1}-\beta_{1,2}=0$
using the regression model $r_{2,1,t} = \alpha_{1} + \beta_{1,1} r_{1,1,t} + \beta_{1,2} r_{1,2,t} + \eta_{1,t}$
or the auxiliary regression
\begin{equation}\label{aux_reg1}
r_{2,1,t}-r_{1,1,t} = \alpha_{1} - \delta_{1} r_{1,1,t} + \beta_{1,2} (r_{1,2,t} -r_{1,1,t})+ \eta_{1,t} \,.
\end{equation}
Testing $H_0^{\alpha_1}$ and $H_0^{\delta_1}$ can also be done using the same regression model. 

Consider the vector $\bx^1_t$ containing the four variables involved in \eqref{aux_reg1}, that is, \linebreak $\bx_t^1=(r_{2,1,t}-  r_{1,1,t},1,r_{1,1,t},r_{1,2,t}- r_{1,1,t})'$ with $\mathbb{V}[\bx_t^1]=\bSigma^1$. Hereafter, we omit the superscript on $\bSigma^1$ for simplicity.

Consider also the following system of nodewize regressions on $\bx_t$, where each variable in the system is regressed on the three other variables (only the first three equations matter for what  follows):

\begin{equation}\label{eq_S}
(S): \begin{cases}
r_{2,1,t}-  r_{1,1,t}=\theta_{1,2}\times 1+\theta_{1,3}\times r_{1,1,t}+\theta_{1,4}\times (r_{1,2,t}- r_{1,1,t})+v_{1,t}\\
1=\theta_{2,1}\times (r_{2,1,t}-  r_{1,1,t})+\theta_{2,3}\times r_{1,1,t}+\theta_{2,4}\times (r_{1,2,t}- r_{1,1,t})+v_{2,t}\\
r_{1,1,t}=\theta_{3,1}\times( r_{2,1,t}-  r_{1,1,t})+\theta_{3,2}\times 1+\theta_{3,4}\times (r_{1,2,t}- r_{1,1,t})+v_{3,t}\\
r_{1,2,t}- r_{1,1,t}=\theta_{4,1}\times (r_{2,1,t}-  r_{1,1,t})+\theta_{4,2}\times 1 +\theta_{4,3}\times r_{1,1,t} +v_{4,t},
\end{cases}
\end{equation} 
with $\mathbb{V}[v_{i,t}]=g^2_i>0$ for $i=1,\ldots,4$,  $\bG\equiv\operatorname{Diag}(g^2_1,\ldots,g^2_4)$ and $\mathbb{E}[x_{i,t}v_{j,t}]=0$ for $i\neq j$ or more compactly 
$\bx^1_t = \bTheta^1\bx^1_t + \bv_t$, where
$\bTheta^1 \equiv \tiny{\left(\begin{array}{cccc}
0 & \theta_{1,2} & \theta_{1,3} & \theta_{1,4} \\
\theta_{2,1} & 0 & \theta_{2,3} & \theta_{2,4} \\
\theta_{3,1} & \theta_{3,2} & 0 & \theta_{3,4} \\
\theta_{4,1} & \theta_{4,2} & \theta_{4,3} & 0
\end{array}\right) }\normalsize$ and $\bv_t\equiv(v_{1,t},\ldots,v_{4,t})'$. 
Therefore, $(\bI_4-\bTheta^1)\mathbb{E}[\bx_t^1\bx_t^{1'}]=\bG$ so that $\bSigma^{-1}=\mathbb{E}[\bx_t^1\bx_t^{1'}]^{- 1}=\bG^{-1}(\bI_4-\bTheta^1)$ if $\mathbb{E}[\bx_t^1\bx_t^{1'}]$ is invertible. 

Comparing the first three equations in \eqref{eq_S} with \eqref{aux_reg1} and the linear models in columns 3 and 4 in Table~\ref{tab:hypo} for $K=2$ and $N=1$, that is, 
$1=b_{1,1} r_{1,1,t}+b_{1,2} r_{1,2,t}+b_{2,1} r_{2,1,t}+u_t$ and 
$r_{1,1,t} = c+ w^{\delta}_2 (r_{1,1,t} -r_{1,2,t}) + w^{\delta}_3 (r_{1,1,t} -r_{2,1,t})+e_t$, we can deduce by identification that

\begin{equation*}\label{symetry}
\bSigma^{-1}=\tiny{\left(\begin{array}{cccc}
1/g_1^2 & -\theta_{1,2}/g_1^2 & -\theta_{1,3}/g_1^2 & -\theta_{1,4}/g_1^2  \\
-\theta_{2,1}/g_2^2 & 1/g_2^2 & -\theta_{2,3}/g_2^2 & -\theta_{2,4}/g_2^2 \\
-\theta_{3,1}/g_3^2 & -\theta_{3,2}/g_3^2  & 1/g_3^2  & -\theta_{3,4}/g_3^2  \\
-\theta_{4,1}/g_4^2  & -\theta_{4,2}/g_4^2 & -\theta_{4,3}/g_4^2 & 1/g_4^2
\end{array}\right)}\normalsize=\tiny{\left(\begin{array}{cccc}
1/g_1^2 & -\alpha_{1}/g_1^2 & \delta_{1}/g_1^2 & -\beta_{1,2}/g_1^2  \\
-b_{2,1}/g_2^2 & 1/g_2^2 & -(b_{1,1}-b_{2,1} - b_{1,2})/g_2^2 & -b_{1,2}/g_2^2 \\
w^\delta_{3}/g_3^2 & -c /g_3^2  & 1/g_3^2  & w^\delta_{2}/g_3^2  \\
-\theta_{4,1}/g_4^2 &-\theta_{4,2}/g_4^2  & -\theta_{4,3}/g_4^2 & 1/g_4^2
\end{array}\right)}.\normalsize
\end{equation*}

Similarly, since $\bv_t=(\bI_4-\bTheta^1)\bx_t^1$, it holds that $\mathbb{E}[\bv_t\bv_t']=(\bI_4-\bTheta^1)\bG$ and that 

\begin{equation*}
\mathbb{E}[\bv_t\bv_t']=\tiny{\left(\begin{array}{cccc}
g_1^2 & -\alpha_{1}g_1^2 & \delta_{1}g_1^2 & -\beta_{1,2}g_1^2  \\
-b_{2,1}g_2^2 & g_2^2 & -(b_{1,1}-b_{2,1} - b_{1,2})g_2^2 & -b_{1,2}g_2^2 \\
w^\delta_{3}g_3^2 & -c g_3^2  & g_3^2  & w^\delta_{2}g_3^2  \\
-\theta_{4,1}g_4^2 &-\theta_{4,2}g_4^2  & -\theta_{4,3}g_4^2 & g_4^2
\end{array}\right)}\normalsize.	
\end{equation*}

We deduce that $\mathbb{E}[v_{1,t} v_{2,t}]=-\alpha_{1}g_1^2$, and that $\mathbb{E}[v_{1,t} v_{3,t}]=\delta_{1}g_1^2$, where $v_{1,t}=\varepsilon(x_{1,t}^1,\bx_{-1,t}^1,\btheta_{1,1})$, $ v_{2,t}=\varepsilon(x_{2,t}^1,\bx_{-2,t}^1,\btheta_{1,2})$ and $v_{3,t}=\varepsilon(x_{3,t}^1,\bx_{-3,t}^1,\btheta_{1,3})$ 
with $\btheta_{1,1}= (\theta_{1,2}, \theta_{1,3}, \theta_{1,4})$,
$\btheta_{1,2}= (\theta_{2,1}, \theta_{2,3}, \theta_{2,4})$
and
$\btheta_{1,3}= (\theta_{3,1}, \theta_{3,2}, \theta_{3,4})$
using the notation of Proposition \ref{th:recast}
so that 
assumptions $H_0^{\alpha_1}$ and $H_0^{\delta_1}$ 
can be tested using  moment conditions, that is, 
$H_0^{\alpha_1}: \mathbb{E}[v_{1,t} v_{2,t}]=0$, $H_0^{\delta_1}: \mathbb{E}[v_{1,t} v_{3,t}]=0$.

We summarize and generalize the above results in Lemma \ref{lemma1} in the online appendix, which implies that Proposition \ref{th:recast} is valid for any $N$ and $K<T$. Let $\hat{\btheta}$ denote the least square estimator of $\btheta$. Section \ref{subsec:CCT} introduces our batch-mean Cauchy combination spanning (BCS) tests using the empirical counterpart of moment conditions defined above. Those empirical moment conditions are 

\begin{itemize}
\item for $H_0^{\balpha}$, $\hat\bm^{\balpha}(\hat\btheta)\equiv\frac{1}{T}\sum_{t=1}^{T}\hat\bg_{\balpha} (\br_{t},\hat\btheta)$, where
\begin{equation*}\label{g_alpha}
\hat\bg_{\balpha} (\br_{t},\hat\btheta)\equiv\left(\hat\varepsilon(x_{1,t}^1,\bx_{-1,t}^1,\hat\btheta_{1,1})\hat\varepsilon(x_{2,t}^1,\bx_{-2,t}^1,\hat\btheta_{1,2}),\ldots,\hat\varepsilon(x_{1,t}^N,\bx_{-1,t}^N,\hat\btheta_{N,1})\hat\varepsilon(x_{2,t}^N,\bx_{-2,t}^N,\hat\btheta_{N,2}) \right)',
\end{equation*}
$\hat\varepsilon(x_{1,t}^j,\bx_{-1,t}^j,\hat\btheta_{j,1})\equiv x_{1,t}^j-\hat\btheta_{j,1}\bx_{-1,t}^j$ with $\hat\btheta_{j,1}\equiv\left(\sum_{t=1}^T\bx_{-1,t}^{j}\bx_{-1,t}^{j'}\right)^{-1}\sum_{t=1}^T\bx_{-1,t}^j x_{1,t}^j \in \mathbb{R}^{K+1}$, for $j=1,\ldots,N$ and $\hat\varepsilon(x_{2,t}^j,\bx_{-2,t}^j,\hat\btheta_{j,2})\equiv x_{2,t}^j-\hat\btheta_{j,2}\bx_{-2,t}^j$ with \linebreak 
$\hat\btheta_{j,2}\equiv\left(\sum_{t=1}^T\bx_{-2,t}^{j}\bx_{-2,t}^{j'}\right)^{-1}\sum_{t=1}^T\bx_{-2,t}^j x_{2,t}^j \in \mathbb{R}^{K+1}$ for $j=1,\ldots,N$. 	
\item for $H_0^{\bdelta}$, $\hat\bm^{\bdelta}(\hat\btheta)\equiv\frac{1}{T}\sum_{t=1}^{T}\hat\bg_{\bdelta} (\br_{t},\hat\btheta)$ where
\begin{equation*}\label{g_delta}
\hat\bg_{\bdelta} (\br_{t},\hat\btheta)\equiv\left(\hat\varepsilon(x_{1,t}^1,\bx_{-1,t}^1,\hat\btheta_{1,1})\hat\varepsilon(x_{3,t}^1,\bx_{-3,t}^1,\hat\btheta_{1,3}),\ldots,\hat\varepsilon(x_{1,t}^N,\bx_{-1,t}^N,\hat\btheta_{N,1})\hat\varepsilon(x_{3,t}^N,\bx_{-3,t}^N,\hat\btheta_{N,3}) \right)',
\end{equation*}
$\hat\varepsilon(x_{3,t}^j,\bx_{-3,t}^j,\hat\btheta_{j,3})\equiv x_{3,t}^j-\hat\btheta_{j,3}\bx_{-3,t}^j$ with $\hat\btheta_{j,3}\equiv\left(\sum_{t=1}^T\bx_{-3,t}^{j}\bx_{-3,t}^{j'}\right)^{-1}\sum_{t=1}^T\bx_{-3,t}^j x_{3,t}^j\in \mathbb{R}^{K+1}$ for $j=1,\ldots,N$.

\item for $H_0^{\balpha,\bdelta}$, $\hat\bm^{\balpha,\bdelta}(\hat\btheta)\equiv\left(\hat\bm^{\balpha}(\hat\btheta),\hat\bm^{\bdelta}(\hat\btheta)\right)' =\frac{1}{T}\sum_{t=1}^{T}\left(\hat\bg_{\balpha} (\br_{t},\hat\btheta)',\hat\bg_{\bdelta} (\br_{t},\hat\btheta)'\right)'$.
\end{itemize}

To avoid heavy notations hereafter, we will use the convention $\btheta_{N+j}=\btheta_j$ and $\hat\btheta_{N+1}=\hat\btheta_j$ for $j=1,\ldots,N$. We will also let $\bm(\btheta)\equiv\bm^{\balpha,\bdelta}(\btheta)$ and 
\begin{align}
	\hat\bm^{\balpha,\bdelta}(\hat\btheta)&\equiv
\hat\bm(\hat{\btheta})\\
	&=\frac{1}{T}\sum_{t=1}^{T}\left(\hat g_{\alpha_1}(\br_{t},\hat\btheta_1),\ldots,\hat g_{\alpha_N}(\br_{t},\hat\btheta_N),\hat g_{\delta_1}(\br_{t},\hat\btheta_{1}),\ldots,\hat g_{\delta_N}(\br_{t},\hat\btheta_{N})\right)'\,,\\
&=\frac{1}{T}\sum_{t=1}^{T}\left(\hat g_1(\br_{t},\hat\btheta_1),\ldots,\hat g_N(\br_{t},\hat\btheta_N),\hat g_{N+1}(\br_{t},\hat\btheta_{N+1}),\ldots,\hat g_{2N}(\br_{t},\hat\btheta_{2N})\right)' \,,\\
&=\frac{1}{T}\sum_{t=1}^{T}\hat\bg (\br_{t},\hat\btheta)\,.
\end{align}

Since the returns are assumed to be covariance-stationary with positive definite covariance matrix, least squares theory ensures that $\hat{\btheta}\overset{p}{\to} \btheta$ and the continuous mapping theorem suggests that  $\bm(\hat{\btheta})\overset{p}{\to} \mathbb{E}\left[\left(\bg_{\balpha} (\br_{t},\btheta)',\bg_{\bdelta} (\br_{t},\btheta)'\right)'\right]$. To be unbiased, the BCS tests requires the following assumptions to be satisfied.

\begin{assumption} \label{A2}  $\hat\btheta \equiv (\hat\btheta_1,\ldots, \hat \btheta_d)'$ is a $\sqrt{T}$-consistent estimator of $\btheta$, the unique vector satisfying $\mathbb{E}[\hat\bm(\btheta)]=\bO_{2N\times 1}$.%
\end{assumption}

\begin{assumption}\label{assump2}  	
$\bg(\br_t,\btheta)$ is the $t$-th observation of a stationary and ergodic process. 
\end{assumption}

Assumption~\ref{A2} ensures the identifiability of $H_0$ and is satisfied as $\hat{\btheta}$ is a least square estimator obtained on stationary data with positive definite covariance matrix. The stationarity and ergodicity conditions in Assumption~\ref{assump2} are also satisfied since $\hat\bg(\br_t,\btheta)$ is defined in Proposition~\ref{th:recast} as a stationary linear combination of $\br_t$ coordinates. 

We now introduce the  BCS tests. Wlog we focus on testing $H_0^{\balpha,\bdelta}:\,\mathbb{E}\left[\left(\bg_{\balpha} (\br_{t},\btheta)',\bg_{\bdelta} (\br_{t},\btheta)'\right)\right]=\bO_{2N\times 1}$.

\subsection{The Batch-Mean Cauchy Combination Spanning (BCS) Tests}\label{subsec:CCT}

Hereafter, we present the BCS tests. Unlike score and Fisher (or Wald) tests, they are reliable and feasible even when $2N \gg T$. As such, the BCS tests can (i) replace the traditional test statistics in high dimensions, or (ii) deal with time-series dependence. We illustrate these properties via an extensive Monte Carlo simulation in Section~\ref{sec:mcsim}.

In the first step of the testing procedure, we perform individual tests $H_{0}^j: \mathbb{E}[m_j( \btheta_j)] = 0$ for $j=1,\ldots,2N$ using a non-overlapping batch-mean method. We recall that by convention $\btheta_{N+j}=\btheta_j$ and $\hat\btheta_{N+1}=\hat\btheta_j$ for $j=1,\ldots,N$.

To do so, we first estimate $\btheta$ over the entire sample of $T$ observations and denote it $\hat{\btheta}$. Next, given the latter, we consider a partition of the observations $\{\br_t\}_{t=1}^T$ into $B$ non-overlapping and consecutive blocks of equivalent sizes $T_b$ $(b=1,\ldots,B)$ such that $T_b/T\to 0$. 
According to \citet{flegal2010batch}, a fixed-width rule $B\equiv [T^\zeta]$, where $[T^\zeta]$ denotes the integer part of $T^\zeta$, can be used to choose the number of blocks $B$. Their analysis suggests that $\zeta$ could be set to $2/3,\; 1/2$, or $1/3$ in practice. Denote $I_b$ the set of indices of the observations belonging to the $b$-th block. We compute

\begin{equation} \label{subvalue}
\hat m_j(\hat\btheta_j )_b
\equiv \frac{1}{T_b} \sum_{t \in I_b} \hat g_j(\br_{t},\hat \btheta_j)\,,
\end{equation} 
for $b=1,\ldots,B$, and

\begin{equation} \label{varsub}
\hat v_{j}(\hat\btheta_j )_{\!B} \equiv \frac{1}{B-1}\sum_{b=1}^B \left(\hat m_j(\hat \btheta_j )_b -\bar m_j(\hat\btheta_j )_{\!B}\right)^2 \,,
\end{equation}
where $\bar m_j(\hat\btheta_j )_{\!B} \equiv \frac{1}{B}\sum_{b=1}^B \hat m_j(\hat\btheta_j )_{b}$. $\hat v_{j}(\hat\btheta_j )_{\!B}$ is called a batch-mean based estimator of $\mathbb V[\hat m_j(\hat\btheta_j)]$. 

\begin{theorem}\label{th:batch}
Assume that Assumptions \ref{A2} and \ref{assump2} hold. Under $H_{0}^j: \mathbb{E}[ m_j(\btheta_j)] = 0$, one has   
\begin{equation}\label{robttest}
t_{j,B}\equiv \sqrt{B} \left( \frac{\bar m_j(\hat \btheta_j )_{\!B}}{\sqrt{\hat v_{j}(\hat\btheta_j )_{\!B} }} \right) \xrightarrow[T\to\infty]{\mathcal{L}} \mathcal{S}t_{B-1} \,,
\end{equation}
where $\hat v_{j}(\hat\btheta_j )_{\!B} - \mathbb{V}[m_j(\btheta_j )] =  o_p(1)$ and $\mathcal{S}t_{\upsilon}$ denotes a Student-t distribution with $\upsilon$ degrees of freedom. The p-value associated to 
$H_{0}^j$ vs. the alternative hypothesis $H_{1}^j: \mathbb{E}[ m_j(\btheta_j)] \neq 0$ is 
$p_{t_{j,B}} = 2\{ 1 - \Phi_{B-1} (| t_{j,B} |)\}$, where $\Phi_{\upsilon} (.)$ is the cumulative distribution fonction of the Student-t distribution with $\upsilon$ degrees of freedom.
\end{theorem}

\begin{remark}\label{rmq:whyBatch}
The standard batch-mean method \citep{carlstein1986use,sherman1997subseries,flegal2010batch,ibragimov2010t,pedersen2020robust} requires refitting the regression above on each batch, that is, the blocks in the batch-mean method. In other words, it requires $K+1<\max_{b=1,\ldots B}T_b $ with $\max_{b=1,\ldots B}T_b\approx T/B$ if the batches have the same length. However, this is not feasible in many empirical applications. For example, when $N=306$, $K=102$, and $T=250$ (see Section \ref{sec:empirics}), $B$ can be at least $6$ when the fixed-width rule of \citet{flegal2010batch} is used. The standard batch-mean approach fails as $K+1=103 \gg \max_{b=1,\ldots B}T_b \approx 40$. In contrast, our test is equivalent to computing over each subsample $b=1,\ldots,B$, the covariances among those models' residuals -- estimated using $\hat\btheta$ -- as we are using the moment conditions in Proposition~\ref{th:recast}. Hence, we never require refitting models and make the batch-mean method applicable even in settings where $K+1>\max_{b=1,\ldots, B}T_b$. Therefore, we reduce the computational burden of the batch-mean method as we only require fitting $N+2$ models of size $K$ instead of $BN$ of size $K$ -- when using the standard batch-mean method. 
\end{remark}

Let $\mathbb{D}\subseteq \{1,\ldots,2N\}$ with $\mathbb{D}$ having cardinality $d>0$. One let $\mathbb{D}=\{1,\ldots,N\}$ (or $\mathbb{D}= \{N+1,\ldots,2N\}$) to test $H_0^{\balpha}$ (or $H_0^{\bdelta}$). To test $H_0^{\balpha,\bdelta}$, one let $\mathbb{D}= \{1,\ldots,2N\}$. In any case, to test the global null hypothesis $H_0=\bigcap_{j\in \mathbb{D}} H_0^j$  without having to compute the covariances between the $2N$ elements $\bar m_j(\hat \btheta_j)_B$, and to control for false discoveries, we use the Cauchy combination test (CCT) introduced by \citet{liu2020cauchy}. The CCT combines the individual p-values of the test statistics of the $d$ null hypotheses $H_0^j$ and provides a joint test for the null hypothesis $H_0$ that accounts for various kinds of dependence between the test statistics (including cross-sectional dependence and serial correlation). The validity of the CCT has been derived under very weak assumptions such as the bivariate normality of each pair of test statistics \citep{liu2020cauchy} or copula arguments \citep{long2022cauchy}. We extend this result and prove that the CCT effectively combines student test statistics.

Let us collect the individual p-values associated with each null hypothesis $H_{0}^j$ defined above. Theorem~\ref{th:cct} introduces a new test of the joint null hypothesis $H_0=\bigcap_{j\in \mathbb{D}} H_0^j$. 

\begin{theorem}~\label{th:cct} 
Let $\bp \equiv (p_1,\ldots,p_d)'$ be a vector of $d$ p-values, where $p_j$ is the p-value 
corresponding to the null hypothesis $H_{0}^j$ for $j\in \mathbb{D}$ and
\begin{equation}\label{ccm}
\textit{CCT}(\bp)\equiv 0.5-\pi^{-1}\arctan{\left[\sum_{j=1}^d \omega_j\tan((0.5-p_j)\pi)\right]}\,,
\end{equation}
be the Cauchy combination test p-value of the joint null hypothesis $H_0=\bigcap_{j\in \mathbb{D}} H_0^j$ with $\bomega\equiv(\omega_1,\ldots,\omega_d)'$ 
a vector of weights independent of $\bp$ with $\sum_{j=1}^{d}\omega_j=1$. Under Assumptions~\ref{A2} and \ref{assump2}, for any $i\neq j$, $(p_i,p_j)'$ forms a bi-variate Student-t copula with $B-1$ degrees of freedom. Then, for small $\alpha^*$, it holds that,
under the global null hypothesis $H_0$,
$\mathbb{P}[\textit{CCT}(\bp)\leq \alpha^*]= \alpha^*$ as $T\to \infty$.
\end{theorem}

Theorem~\ref{th:cct} follows from the fact that the distribution of each pair $(p_i,p_j)$ for $i\neq j$ can be represented using a Student-t copula with $B-1$ degrees of freedom. In fact, we deduce from \citet[Equation 13]{demarta2005t} that $p_i$ and $p_j$ are asymptotically tail independent when $T\to \infty$ as $B=[T^\zeta]\to \infty$ for $\zeta>0$. That is, we also deduce from \citet{chen2009sums} that
\begin{align}\label{heavytest}
\lim_{\tau\to +\infty}\frac{\mathbb{P}\left[  \sum_{j=1}^d \omega_j\tan((0.5-p_j)\pi) >\tau \right]}{\mathbb{P}[C>\tau]} = 1 \,,
\end{align}
where $C$ follows a standard Cauchy random variable with $\bomega$ independent of $\bp$.  

\begin{remark} 
When $B\to \infty$, these p-values behave asymptotically as if drawn from a Gaussian copula. In that case, we obtain the same result as in \citet{liu2020cauchy}, and we also deduce from \citet[Theorem 5]{long2022cauchy} that (spanning tests based on) the $\textit{CCT}$ has power no less than that of the supremum test when $T\to \infty$. Importantly, the tail independence between $p_i$ and $p_j$ for $i\neq j$ also holds when p-values are correlated and when $B$ is relatively small \citep[see, \eg,][Table~1]{demarta2005t}. This is very important because in finite samples, $B=[T^\zeta]$ can be pretty small for some $\zeta>0$. \citet{liu2020cauchy} also show via Monte Carlo simulations that the CCT has good finite sample properties when the test statistics follow a multivariate Student-t distribution with four degrees of freedom. According to \citet[Theorem 3.2]{ling2023additive}, whenever $p_i$ and $p_j$ are tail independent for all $i\neq j$, the $\textit{CCT}$ has asymptotic optimal power for large $d$. 
\end{remark} 

It is worth mentioning that the CCT is an alternative to popular multiple testing corrections, such as statistical inequalities, including the so-called Bonferroni correction and its subsequent improvements \citep{holm1979simple,hommel1988stagewise,hochberg1988sharper} that are known to be overly conservative or those based on the extreme value theory, and in particular, the Gumbel distribution \citep[\eg,][]{lee2007jumps} that assumes the test-statistics are \textit{i.i.d.} under the null hypothesis, which is overly restrictive.

\subsection{Practical Implementation}
\label{sec:practical}

The idea of the batch-mean method is to eliminate the dependence between the subseries values (the statistics computed on the blocks) by calculating them on sufficiently large subsamples. The more dependence there is, the more observations we need to take. But the smaller the number of blocks, the less powerful and size-distorted the test is because there is not enough variability to perform a good t-test---indeed, a decent number of subseries values shall be used to calculate the variance of their mean. This trade-off is limiting as the sample size is finite. Therefore, we propose to use a randomly-weighted batch-mean method. This method allows us to break the dependence between the subseries values artificially and between the observations that generate these subseries values. By breaking the covariation between the observations over time, we can take a reasonable number of blocks without being stressed by the strength of the dependence between the variables.

More specifically, we apply the batch-mean method on randomly-weighted empirical moments, that is: 
\begin{equation}\label{subvalue2}
\tilde m_j(\hat\btheta_j )_{b} \equiv \frac{1}{T_b} \sum_{t \in I_b} 
\hat g_j(\br_t,\hat\btheta_j )\kappa_{t},\; \forall\; b=1,\ldots,B \,,
\end{equation} 
instead of \eqref{subvalue}, with $\kappa_t \equiv \prod_{l=1}^L \kappa_{l,t}$ and $\kappa_{l,t}\sim \mathcal{N}(1,1)$ and where we set $\kappa_t \equiv 1$ if $L=0$. Again, the intuition behind this approach is that using $\kappa_t$ in \eqref{subvalue2} kills the dependence in $\hat g_j(\br_t,\hat\btheta_j)$. In fact, under Assumption \ref{assump2} and the global null hypothesis $H_0$, 
$\tilde m_j(\hat\btheta_j)_{b}$ has mean 0 and variance $\mathbb{V}[\tilde m_j(\hat \btheta_j)_b]= \mathbb{V}[\hat g_j(\br_t,\hat\btheta_j)] / T_b$ as $\mathbb{C}ov[\hat g_j(\br_t,\hat\btheta_j)\kappa_{t},\hat g_j(\br_{t'},\hat\btheta_j)\kappa_{t'} ]=0$ whenever $t\neq t'$. Moreover, if Assumptions \ref{A2} and \ref{assump2} hold for $g_j(\br_t,\hat\btheta_j)\kappa_{t}$, Theorem \ref{th:batch} holds when $\hat g_j(\br_t,\hat\btheta_j)\kappa_{t}$ replaces $\hat g_j(\br_t,\hat\btheta_j)$.

The procedure is equivalent to a mixture of the multiplier bootstrap \citep{zhang2017gaussian} and the batch-mean method \citep{carlstein1986asymptotic}. Our combination of both methods reduces the computational cost of the bootstrap as it requires only one resampling. However, it preserves the intrinsic features of the batch-mean procedure introduced above, as the latter is then used in a step to approximate the distribution of the test statistic.

We now denote our test as BCS$^{\blambda}_L$ when testing $H_0^{\blambda}$, for $\blambda\in\{\balpha,\bdelta,\{\balpha,\bdelta\}\}$. 
When $L=0$, the p-value of $H_0^{\blambda}$ is obtained by using \eqref{ccm}, where the individual p-values $p_{t_{j,B}}$ depend on the statistics $t_{j,B}$ as described in \eqref{robttest}. When $L>0$, the statistics $t_{j,B}$ are obtained using 
the randomly-weighted versions of \eqref{subvalue} and \eqref{varsub}.
In the Monte Carlo simulation, we will consider three choices of $L$, that is, $L=0$ and 2. Following \citet{liu2020cauchy}, we set $\omega_j=1/d$ in \eqref{ccm}.

\section{Simulation Study}
\label{sec:mcsim}

This section benchmarks our newly proposed MVS tests with the state-of-the-art methods proposed in the literature. We first present the alternative tests and then introduce the data-generating processes used in the simulation study. We finally present the simulation results.

\subsection{Competing Spanning Tests}

Recall \citet{huberman1987mean}'s linear regression model introduced in Table~\ref{tab:hypo}. Let $\bY=\bX \bB + \bE$, where $\bY$ is a $T \times N$ matrix whose $t$-th row is $\br_{2,t}, \bX$ is a $T \times(K+1)$ matrix whose $t$-th  row is $(1, \br_{1,t}'), \bB \equiv [\balpha, \bbeta]'$, and $\bE$ is a $T \times N$ matrix whose $t$-th row is $\etab_t^{\prime}$. Assume that $T \geq N+K+1$, $\bX^{\prime} \bX$ is nonsingular, and that conditional on $\br_{1,t}$, the disturbances $\etab_t$ are \textit{i.i.d.} multivariate normal with zero mean and positive definite covariance matrix. 

\subsubsection{Testing $H_0^{\balpha,\bdelta}$}

\paragraph{HK Test} Define $\hat a \equiv \hat\bmu' \hat\bV^{-1} \hat \bmu, \hat b \equiv \hat\bmu' \hat\bV^{-1} \bi_{N+K}, \hat c \equiv \bi_{N+K}' \hat\bV^{-1} \bi_{N+K}$, $\hat d \equiv \hat a\hat c -\hat b^2$, $a_1 \equiv \hat\bmu_1' \hat\bV_{11}^{-1} \hat\bmu_1$, $b_1 \equiv \hat\bmu_1' \hat\bV_{11}^{-1} \bi_K$, $c_1 \equiv \bi_K' \hat\bV_{11}^{-1} \bi_K$, and $\hat d_1 \equiv \hat a_1\hat c_1 -\hat b_1^2$, where $\hat{\bmu} \equiv \frac{1}{T} \sum_{t=1}^T \br_{t}$, $\hat{\bV} \equiv \frac{1}{T} \sum_{t=1}^T\left(\br_{t}-\hat{\bmu} \right)\left(\br_{t}-\hat{\bmu} \right)'$, $\hat{\bmu}_1 \equiv \frac{1}{T} \sum_{t=1}^T \br_{1,t}$, and $\hat{\bV}_{11} \equiv \frac{1}{T} \sum_{t=1}^T\left(\br_{1,t}-\hat{\bmu}_1\right)\left(\br_{1,t}-\hat{\bmu}_1\right)'$.  Under the normality assumption, \citet{huberman1987mean} define the following $F$-test statistic to test the MVS hypothesis $H_0^{\alpha,\delta}$,
\begin{equation}
\textit{HK} \equiv 
\begin{cases}
\left(\frac{1}{U^{\frac{1}{2}}}-1\right)\left(\frac{T-K-N}{N}\right) \sim F_{2 N, 2(T-K-N)} & \text{if} \quad N \geq 2  \\
\left(\frac{1}{U}-1\right)\left(\frac{T-K-1}{2}\right) \sim F_{2, T-K-1}  
& \text{if} \quad N=1, 
\end{cases} 
\end{equation}
where $U \equiv \frac{\hat c_1+\hat d_1}{\hat c+\hat d}$. Note that $\textit{HK}$ is not applicable when $T-K-N<1$. 

\paragraph{GL Test} \citet{gungor2016multivariate} show that, for $K<T$ and $N>1$, $F_{\max}\equiv\max_{i=1,\ldots N}F_i$ where $F_i$ denotes the $\textit{HK}$ test statistic applied to the $i$th equation in \eqref{eq_tests}. The distribution of $F_{\max}$ is obtained via simulations assuming that the data is stationary and that conditional on $\bX$, $\etab_t$ has the same distribution as $-\etab_t$, that is, the reflexive symmetric assumption; see \citet[Section 3.1]{gungor2016multivariate} for details. The authors argue that their testing procedure can be used to test linear restrictions in any multivariate linear regression model when $K<T$. However, we find that their testing procedure can be non-informative when $K$ is moderately large but still lower than $T$.

\paragraph{BCS$^{\balpha,\bdelta}$ Tests} We also consider two BCS tests, namely BCS$^{\balpha,\bdelta}_0$
and BCS$^{\balpha,\bdelta}_2$. 

\subsubsection{Testing $H_0^{\balpha}$}

\paragraph{GRS}  Recall Model \eqref{eq_tests}. Denote  $\hat\bbeta^c$ the constrained OLS estimator of $\bbeta$ when $\balpha=\bO_N$ is imposed. Denote $(\hat\balpha',\hat\bbeta')'$ the standard OLS estimator of $\left(\balpha,\bbeta\right)'$. The GRS test statistic for $H_0^{\balpha}$ is
\begin{equation}
\textit{GRS} \equiv \frac{(T-N-K)}{N}
\left(\frac{|\hat{\bGamma}_1|}{|\hat{\bGamma}_2|}-1\right),  
\end{equation}
where $$\hat{\boldsymbol{\bGamma}}_1 \equiv \frac{1}{T} \sum_{t=1}^T\left(\br_{2,t}-\hat\bbeta^{c'}\br_{1,t} \right)\left(\br_{2,t}-\hat\bbeta^{c'}\br_{1,t} \right)'\,,
$$ and $$\hat{\boldsymbol{\bGamma}}_2 \equiv \frac{1}{T} \sum_{t=1}^T\left(\br_{2,t}-\hat\balpha-\hat\bbeta'\br_{1,t} \right)\left(\br_{2,t}-\hat\balpha-\hat\bbeta'\br_{1,t} \right)'.
$$
\citet{gibbons1989test} show that \textit{GRS} follows a $F_{N, T-N-K}$ under $H_0^{\balpha}$. 

\paragraph{$F_1$ Test} To test $H_0^{\balpha}$, \citet{kan2012tests} introduce the following F-statistic (that is similar to \citealp{gibbons1989test}'s statistic) 
\begin{equation}
F_1 \equiv \left(\frac{T-K-N}{N}\right)\left(\frac{\hat{a}-\hat{a}_1}{1+\hat{a}_1}\right)\sim F_{N, T-K-N} \,.
\end{equation}

\paragraph{BJ Test} Recall \citet{britten1999sampling} model from Table~\ref{tab:hypo}. These authors also show that
\begin{equation}
\textit{BJ} \equiv \frac{T-N-K}{N}\frac{\left(\mathrm{SSR}_R-\mathrm{SSR}_u\right)}{\mathrm{SSR}_u}\sim F_{N, T-N-K}
\end{equation}
under $H_0^{\balpha}$,  where $\mathrm{SSR}_u$ is the sum of squared residuals of the complete model and $\mathrm{SSR}_R$ is the sum of squared residuals of the model under the restriction that $\bb_1=\bO_{K \times N}$. 

\paragraph{PY Test} \citet{PesaranYamagata2012} show, under some regularity conditions, that for Gaussian and non-Gaussian disturbances $\etab_t$, if $N$ grows at a sufficiently slower rate than $T$, 
\begin{equation}
\textit{PY}  \equiv \frac{N^{-1 / 2} \sum_{i=1}^N\left(t_i^2-\frac{v}{v-2}\right)}{\left(\frac{v}{v-2}\right) \sqrt{\frac{2(v-1)}{v-4}\left[1+(N-1) \hat{\rho}^2\right]}}
\xrightarrow[T\to\infty]{\mathcal{L}} \mathcal{N}(0,1)\,,
\end{equation}
where $t_i$ is the Student-t statistic of the significance of $\alpha_i$ in \eqref{eq_tests}, $v\equiv T-K-1$, and $\hat{\rho}^2$ is a threshold estimator of the average squares of pairwise disturbance correlations given by $\hat{\rho}^2 \equiv \frac{2}{N(N-1)} \sum_{i=2}^N \sum_{j=1}^{i-1} \hat{\rho}_{i j}^2 \mathbb{I}\left[v \hat{\rho}_{i j}^2 \geq \theta_N\right]$,
where $\mathbb{I}$ denotes the indicator function, $\hat{\rho}_{i j} \equiv \hat g_i / \sqrt{\hat g_i \hat g_j}$ and $\hat g_i$ is the $i$-th element on the diagonal of $\hat \bG \equiv \sum_{t=1}^{T}\hat{\etab}_t \hat{\etab}_t'$; recall that $\hat{\etab}_t$ are the OLS residuals from \eqref{eq_tests}.  \citet{PesaranYamagata2012} suggest selecting the threshold value as $\sqrt{\theta_N} \equiv \Phi^{-1}\left(1-\frac{\alpha}{2(N-1)}\right)$, where $\Phi^{-1}(\cdot)$ is the standard normal quantile function. However, \citet{gungor2016multivariate} found that the test is misleading for general correlation structures. In this paper, we show that the test is not robust to serial dependence in general.

\paragraph{BCS$^{\balpha}$ Tests} We also consider two BCS tests, namely BCS$^{\balpha}_0$
and BCS$^{\balpha}_2$. 

\subsubsection{Testing $H_0^{\bdelta}$}

\paragraph{KM Test} Let $S S R$ denote the sum of the squared residuals in the unrestricted model  $r_{1,1,t}=c+\sum_{j=2}^{N+K} w_j(r_{1,1,t}-r_{j,t})+e_t$ and $S S R_r$ be the sum of the squared residuals in the restricted regression $r_{1,1,t}=c^r+\sum_{j=2}^{K}w^r_j (r_{1,1,t}-r_{j,t})+e_t^r$. Let $m \leq N-1$ be the number of linear independent restrictions.  \citet{kempf2006estimating} show that for Gaussian disturbance $e_t$ 
$$
F\equiv\frac{T-N-K}{N}\left(\frac{S S R_r}{S S R}-1\right) \sim F_{N, T-N-K} \,.
$$
\paragraph{F2 Test} \citet{kan2012tests} shows that for Gaussian disturbance $\etab_t$
$$
\begin{aligned}
F_2& \equiv\left(\frac{T-K-N+1}{N}\right)\left[\left(\frac{\hat{c}+\hat{d}}{\hat{c}_1+\hat{d}_1}\right)\left(\frac{1+\hat{a}_1}{1+\hat{a}}\right)-1\right]\sim F_{N,T-K-N+1}.
\end{aligned}
$$

\paragraph{BCS$^{\bdelta}$ Tests} We also consider two BCS tests, namely BCS$^{\bdelta}_0$
and BCS$^{\bdelta}_2$. 

\subsection{The Setup}

Let us assume that the benchmark and test assets are generated by a DGP belonging to the class of  \citet{darolles2018asymptotics}'s Cholesky-GARCH process augmented with an AR component. Inspired by \citet{pedersen2020robust}, we use univariate AR-GARCH processes to reproduce well-known stylized facts on stock returns. Namely, we let
\begin{equation}
\br_{1,t}=\phi \br_{1,t-1}+\bR^{1/2}_{\bar{\rho}_1,1}\underbrace{ \bD^{1/2}_{1,t}\bnu_{1,t}}_{\bG_{1,t}}\label{DGP_eq_first}
\end{equation}
and 
\begin{align}
\br_{2,t}&=\balpha+\bbeta \br_{1,t}+ \etab_t,\\
\etab_t&=\phi\etab_{t-1}+ \bR^{1/2}_{\bar{\rho}_2,2} \underbrace{\bD^{1/2}_{2,t} \bnu_{2,t}}_{\bG_{2,t}},
\label{DGP_eq_last}
\end{align}
where $\phi \in \{0,0.2\}$, $\beta_{i,j}=1$ for $j=2,\ldots,K$, 
$\bG_{1,t}\equiv(g_{1,j,t})$, $\bG_{2,t}\equiv(g_{2,i,t})$, 
$\bD^{1/2}_{1,t}\equiv(d_{1,j,t})$ and  $\bD^{1/2}_{2,t}\equiv(d_{2,i,t})$ are two diagonal matrices of size $K$ and $N$ respectively. $\mathbf{R}_{\bar\rho_l}^{1/2}$ denotes the Cholesky factor of $\mathbf{R}_{\bar\rho_l}\equiv\left(\bar\rho_l^{|i-j|}\right)$, a Toeplitz correlation matrix with coefficient $\bar\rho_l=0.8$ for $l=1$ and $\bar\rho_l=0.5$ for $l=2$. For the specifications with GARCH effects, $d_{1,j,t} \equiv (0.1+ 0.1  g_{1,j,t-1}^2 + 0.8 d_{1,j,t-1}^2)^{1/2}$ for $j=1,\ldots,K$ and $d_{2,i,t} \equiv (0.1+ 0.1  g_{2,i,t-1}^2 + 0.8 d_{2,i,t-1}^2)^{1/2}$ for $i=1,\ldots,N$
so that the elements of $\bG_{1,t}$ and $\bG_{2,t}$ follow GARCH(1,1) specifications. In absence of GARCH effects, $d_{1,j,t}=d_{2,i,t}=1$
for any $i$ and $j$ so that the elements of $\bG_{1,t}$ and $\bG_{2,t}$ are homoscedastic. 
The innovations $\bnu_{1,t}\equiv (\nu_{1,t},\ldots,\nu_{K,t})'$ and $\bnu_{2,t} \equiv (\nu_{K+1,t},\ldots,\nu_{K+N,t})'$ are $i.i.d.$ random variables with mean 0 and will be defined below. Thus, both benchmark and test assets can display serial correlation, correlated innovations, and GARCH effects. 

We use this general DGP to compare the finite sample properties of the F, PY, and GL tests to the BCS tests and later present the benefit of using the randomly-weighted batch-mean method. To identify the possible cause of the underperformance of these tests, we simulate nine specific DGPs corresponding to small variations of the general DGP \eqref{DGP_eq_first}-\eqref{DGP_eq_last}.

\begin{itemize}
\item DGP1 ($i.i.d.$ N). $\phi=0$, $d_{1,j,t}=d_{2,i,t}=1$, $\bnu_{1,t} \overset{i.i.d.}{\sim} \mathcal N(\mathbf{0},I_K)$ and $\bnu_{2,t} \overset{i.i.d.}{\sim} N(\mathbf{0},I_N)$. 
\item DGP2 ($i.i.d$ ST).  Same as DGP1, but the elements of the innovations
$\bnu_{1,t}$ and $\bnu_{2,t}$ follow independent standardized Student-t distributions with $5$ degrees of freedom.  
\item DGP3 ($i.i.d$ SKST). Same as DGP1, but the elements of the innovations 
$\bnu_{1,t}$ and $\bnu_{2,t}$ follow independent standardized skewed Student-t distribution 
originally proposed by \citet{FernandezSteel1998} and modified by  
\citet{giot2003value} to have mean zero and unit variance, with asymmetry parameter $0.9$ and  $4$ degrees of freedom.  
\item DGP4 (GARCH N), DGP5 (GARCH ST) and DGP6 (GARCH SKST). Respectively the same as DGP1, DGP2 and DGP3  but $\mathbf{G}_{1,t}$ and $\mathbf{G}_{2,t}$ have GARCH effects.
\item DGP7 (AR-N), DGP8 (AR-ST) and DGP9 (AR-SKST). Respectively the same as DGP1, DGP2 and DGP3 but $\phi=0.2$ so that $\br_{1,t}$ and $\br_{2,t}$ have serial correlation.
\item DGP10 (AR-GARCH N), DGP11 (AR-GARCH ST), DGP12 (AR-GARCH SKST). Respectively the same as DGP4, DGP5 and DGP6 but $\phi=0.2$ so that $\br_{1,t}$ and $\br_{2,t}$ have serial correlation 
and have GARCH effects.
\end{itemize}

In the next sections, we study the size and power of various tests for the null hypotheses $H_0^{\balpha,\bdelta}$, $H_0^{\balpha}$, and $H_0^{\bdelta}$. The simulation is performed $500$ times for a sample of $T=250$ observations and various values of $N$ and $K$.

\subsection{Size Results}\label{sec:size}

To study the size of the tests, we simulate data under $H_0^{\balpha,\bdelta}$ by setting first $\balpha=\bO_N$ in \eqref{DGP_eq_last}. Recall that $\bdelta\equiv\bi_N-\bbeta \bi_K$ 
and note that $r_{K+i,t}= r_{1,t}+ \alpha_i+\delta_i r_{1,t}+\sum_{j=2}^{K}\beta_{i,j}(r_{j,t}-r_{1,t})+\eta_{i,t}$, $\forall i=1,\ldots,N$. Therefore, we have $\beta_{i,1}=1 - \delta_i - \sum_{j=2}^K \beta_{i,j}$. To set $\bdelta=\bO_N$, we set $\beta_{i,1}=1 - \sum_{j=2}^K \beta_{i,j}$ for $i=1,\ldots,N$, where $\beta_{i,j}=1$ for $j>1$.

\subsubsection{Size Results for $H_0^{\balpha,\bdelta}$}

The empirical size of the HZ, GL and our BCS$^{\balpha,\bdelta}_L$ test with $L=0$ and 2 for the null hypothesis $H_0^{\balpha,\bdelta}$ is reported in Table~\ref{tab:sizeH0alphadelta} for DGP1--DGP12.
The nominal size of each test is 5\%. For the readability of the results, empirical sizes between 3\% and 7\% are highlighted in bold.  
Several comments are in order.
\begin{itemize}
\item HK is robust to fat-tailed and even skewed innovations but is largely oversized in the presence of serial correlation, with rejection frequencies sometimes approaching 70\% when $N$ is 100 and $K$ is small. Furthermore, the test is not applicable when $N=400$, for a sample size of $T=250$. 
\item GL is largely undersized for most configurations, even in the $i.i.d.$ Gaussian case. 
\item When $N$ and $K$ are relatively small, our proposed test BCS$^{\balpha,\bdelta}_0$ has a decent empirical size but is oversized when $N$ and $K$ are greater than 10 for most DGPs. This is especially true in the presence of serial correlation. 
\item Interestingly, the 
BCS$^{\balpha,\bdelta}_2$ test improves considerably. Indeed, the empirical sizes are close to 5\% for most DGPs, even for very large values of $N$ and $K$.
\end{itemize}

Overall, the test with the best empirical size is BCS$^{\balpha,\bdelta}_2$. 

\begin{table}[H]
\centering
\caption{\textbf{Empirical Size Results for $H_0^{\balpha,\bdelta}$}\\
The table reports the empirical size (over 500 replications) for five tests for 
the global null hypothesis of MVS under twelve 
DGPs and for four values of $K$ and $N$ (\ie, the number of benchmark and test assets, respectively). Values in bold are between 3\% and 7\%. A dash indicates that 
the test cannot be applied as $N>T-K-1$.}
\singlespacing
\scalebox{0.6}{
\begin{tabular}{p{0.1cm}p{0.1cm}llccccc|ccccc|ccccc|ccccc|}
\toprule\label{tab:sizeH0alphadelta}
&&&&\multicolumn{5}{c}{HK}
&\multicolumn{5}{c}{GL}
&\multicolumn{5}{c}{BCS$^{\balpha,\bdelta}_0$}
&\multicolumn{5}{c}{BCS$^{\balpha,\bdelta}_2$}\\
\cmidrule(lr){5-9}
\cmidrule(lr){10-14}
\cmidrule(lr){15-19}
\cmidrule(lr){20-24}
&&&$K\!\!\downarrow N\!\!\rightarrow$&  2 & 10 & 50 & 100 & 400 & 2 & 10 & 50 & 100 & 400 & 2 & 10 & 50 & 100 & 400 & 2 & 10 & 50 & 100 & 400\\ 
\midrule
\multirow{4}{*}{\rotatebox[origin=c]{90}{DGP1}}&\multirow{4}{*}{\rotatebox[origin=c]{90}{$i.i.d$}}& \multirow{4}{*}{\rotatebox[origin=c]{90}{N}}&2 & $\grbb{5.2}$ & $\grbb{4.8}$ & $\grbb{5.2}$ & $\grbb{6.2}$ & - & 1.8 & 1.4 & 0.6 & 0.6 & 1.0 & $\grbb{6.0}$ & $\grbb{5.0}$ & $\grbb{6.8}$ & $\grbb{5.8}$ & $\grbb{5.4}$ & $\grbb{4.8}$ & $\grbb{6.0}$ & $\grbb{5.0}$ & 2.6 & $\grbb{4.2}$ \\ 
&&&10  & $\grbb{6.2}$ & $\grbb{5.6}$ & $\grbb{3.8}$ & $\grbb{5.8}$ & - & 0.0 & 0.0 & 0.0 & 0.0 & 0.0 & $\grbb{5.4}$ & $\grbb{5.2}$ & $\grbb{6.2}$ & $\grbb{6.6}$ & 7.4 & $\grbb{5.8}$ & $\grbb{4.4}$ & $\grbb{3.6}$ & $\grbb{5.4}$ & $\grbb{3.0}$ \\ 
&&&50 & $\grbb{6.2}$ & $\grbb{6.8}$ & $\grbb{5.6}$ & $\grbb{6.0}$ & - & 0.0 & 0.0 & 0.0 & 0.0 & 0.0 & $\grbb{6.4}$ & 10.0 & 8.0 & 10.0 & 10.8 & $\grbb{5.2}$ & $\grbb{3.0}$ & $\grbb{4.2}$ & $\grbb{4.2}$ & $\grbb{3.2}$ \\ 
&&&100 & $\grbb{5.0}$ & $\grbb{4.0}$ & $\grbb{5.2}$ & $\grbb{6.4}$ & - & 0.0 & 0.0 & 0.0 & 0.0 & 0.0 & 8.0 & 16.0 & 18.6 & 21.4 & 25.8 & $\grbb{5.8}$ & $\grbb{4.4}$ & $\grbb{3.8}$ & $\grbb{6.4}$ & $\grbb{6.4}$ \\ 
\midrule
\multirow{4}{*}{\rotatebox[origin=c]{90}{DGP2}}&\multirow{4}{*}{\rotatebox[origin=c]{90}{$i.i.d$}}& \multirow{4}{*}{\rotatebox[origin=c]{90}{ST}}&2 & $\grbb{4.2}$ & $\grbb{3.4}$ & $\grbb{5.4}$ & $\grbb{5.0}$ & - & 2.6 & 1.2 & 2.6 & 2.0 & 2.2 & $\grbb{5.4}$ & 7.2 & 7.2 & $\grbb{5.2}$ & $\grbb{5.2}$ & $\grbb{4.0}$ & $\grbb{4.4}$ & $\grbb{3.6}$ & 2.6 & $\grbb{3.2}$ \\ 
&&&10 & $\grbb{4.0}$ & $\grbb{5.0}$ & $\grbb{3.4}$ & $\grbb{4.6}$ & - & 0.0 & 0.0 & 0.0 & 0.0 & 0.0 & $\grbb{4.6}$ & $\grbb{6.2}$ & $\grbb{6.6}$ & $\grbb{6.8}$ & $\grbb{7.0}$ & $\grbb{5.2}$ & $\grbb{3.6}$ & $\grbb{4.0}$ & $\grbb{4.6}$ & $\grbb{4.0}$ \\ 
&&&50 & $\grbb{4.0}$ & $\grbb{5.8}$ & $\grbb{5.0}$ & $\grbb{5.2}$ & - & 0.0 & 0.0 & 0.0 & 0.0 & 0.0 & 8.8 & 11.2 & 10.4 &  9.8 & 11.4 & $\grbb{5.2}$ & $\grbb{3.8}$ & $\grbb{5.2}$ & $\grbb{4.6}$ & $\grbb{3.4}$ \\ 
&&&100 & $\grbb{5.6}$ & $\grbb{5.6}$ & $\grbb{4.6}$ & $\grbb{6.6}$ & - & 0.0 & 0.0 & 0.0 & 0.0 & 0.0 & 15.4 & 19.8 & 17.8 & 21.6 & 27.0 & $\grbb{5.0}$ & $\grbb{6.2}$ & $\grbb{5.6}$ & $\grbb{5.6}$ & $\grbb{5.4}$ \\ 
\midrule
\multirow{4}{*}{\rotatebox[origin=c]{90}{DGP3}}&\multirow{4}{*}{\rotatebox[origin=c]{90}{$i.i.d$}}& \multirow{4}{*}{\rotatebox[origin=c]{90}{SKST}}&2 & $\grbb{5.6}$ & $\grbb{5.8}$ & $\grbb{3.4}$ & $\grbb{5.8}$ & - & 1.6 & 2.2 & 0.8 & 1.2 & 1.0 & $\grbb{6.8}$ & $\grbb{5.4}$ & $\grbb{4.0}$ & $\grbb{4.8}$ & 7.4 & $\grbb{4.0}$ & 2.6 & $\grbb{4.0}$ & $\grbb{3.4}$ & 2.8 \\ 
&&&10& $\grbb{5.2}$ & $\grbb{4.6}$ & $\grbb{6.4}$ & $\grbb{5.2}$ & - & 0.0 & 0.0 & 0.0 & 0.0 & 0.0 & 7.4 & $\grbb{6.6}$ & $\grbb{5.0}$ & $\grbb{5.8}$ & $\grbb{5.8}$ & $\grbb{5.8}$ & $\grbb{3.6}$ & 2.4 & 2.0 & $\grbb{3.4}$ \\ 
&&&50 & $\grbb{5.6}$ & $\grbb{7.0}$ & $\grbb{6.0}$ & $\grbb{3.6}$ & - & 0.0 & 0.0 & 0.0 & 0.0 & 0.0 & 8.0 & 11.4 & 12.8 & 13.4 & 12.0 & $\grbb{5.0}$ & $\grbb{3.8}$ & $\grbb{5.6}$ & $\grbb{4.2}$ & $\grbb{4.2}$ \\ 
&&&100 & $\grbb{4.6}$ & $\grbb{3.4}$ & $\grbb{4.0}$ & $\grbb{4.2}$ & - & 0.0 & 0.0 & 0.0 & 0.0 & 0.0 & 15.2 & 16.0 & 23.2 & 24.0 & 31.2 & 7.2 & $\grbb{4.2}$ & $\grbb{5.0}$ & $\grbb{5.0}$ & $\grbb{5.2}$ \\
\midrule\midrule
\multirow{4}{*}{\rotatebox[origin=c]{90}{DGP4}}&\multirow{4}{*}{\rotatebox[origin=c]{90}{GARCH}}& \multirow{4}{*}{\rotatebox[origin=c]{90}{N}}&2 & $\grbb{5.2}$ & $\grbb{5.0}$ & $\grbb{4.8}$ & $\grbb{5.0}$ & - & 2.2 & 1.0 & 1.2 & 1.4 & 1.2 & $\grbb{4.8}$ & $\grbb{3.8}$ & $\grbb{5.0}$ & $\grbb{5.0}$ & $\grbb{5.6}$ & $\grbb{4.4}$ & $\grbb{4.4}$ & $\grbb{3.6}$ & $\grbb{3.6}$ & 2.8 \\ 
&&&10 & $\grbb{4.4}$ & $\grbb{3.6}$ & $\grbb{4.0}$ & $\grbb{5.6}$ & - & 0.0 & 0.0 & 0.0 & 0.0 & 0.0 & $\grbb{4.8}$ & $\grbb{4.8}$ & $\grbb{6.0}$ & $\grbb{4.4}$ & $\grbb{4.8}$ & $\grbb{4.0}$ & $\grbb{3.0}$ & $\grbb{3.0}$ & $\grbb{4.8}$ & $\grbb{4.8}$ \\ 
&&&50  & $\grbb{4.8}$ & $\grbb{5.6}$ & $\grbb{3.8}$ & $\grbb{7.0}$ & - & 0.0 & 0.0 & 0.0 & 0.0 & 0.0 & $\grbb{6.8}$ & 8.6 & 10.0 & 9.0 & 8.6 & $\grbb{4.2}$ & $\grbb{6.6}$ & $\grbb{5.8}$ & $\grbb{4.8}$ & $\grbb{4.6}$ \\ 
&&&100  & $\grbb{5.4}$ & $\grbb{4.8}$ & $\grbb{6.8}$ & 7.2 & - & 0.0 & 0.0 & 0.0 & 0.0 & 0.0 & 14.2 & 17.0 & 24.2 & 24.0 & 34.2 & $\grbb{5.0}$ & $\grbb{6.0}$ & $\grbb{5.2}$ & $\grbb{4.0}$ & $\grbb{7.0}$ \\
\midrule
\multirow{4}{*}{\rotatebox[origin=c]{90}{DGP5}}&\multirow{4}{*}{\rotatebox[origin=c]{90}{GARCH}}& \multirow{4}{*}{\rotatebox[origin=c]{90}{ST}}&2 & $\grbb{3.6}$ & $\grbb{4.8}$ & $\grbb{5.0}$ & $\grbb{6.4}$ & - & 1.6 & 1.6 & 1.6 & 2.0 & 1.4 & 9.4 & 8.0 & 8.6 & 8.4 & $\grbb{6.4}$ & $\grbb{4.0}$ & $\grbb{4.4}$ & $\grbb{5.0}$ & $\grbb{3.4}$ & 2.8 \\ 
&&&10  & $\grbb{5.2}$ & $\grbb{4.4}$ & $\grbb{5.0}$ & $\grbb{5.2}$ & - & 0.0 & 0.0 & 0.0 & 0.0 & 0.0 & $\grbb{6.8}$ & 8.4 & 8.2 & 9.2 & $\grbb{5.8}$ & $\grbb{4.0}$ & $\grbb{5.2}$ & 2.6 & $\grbb{5.4}$ & 2.4 \\ 
&&&50  & $\grbb{3.8}$ & $\grbb{4.2}$ & $\grbb{5.2}$ & $\grbb{6.2}$ & - & 0.0 & 0.0 & 0.0 & 0.0 & 0.0 & 7.6 & 12.6 & 14.4 & 13.2 & 12.2 & $\grbb{5.4}$ & $\grbb{4.2}$ & $\grbb{5.6}$ & $\grbb{4.0}$ & $\grbb{4.2}$ \\ 
&&&100  & $\grbb{5.0}$ & $\grbb{4.6}$ & $\grbb{4.4}$ & $\grbb{6.0}$ & - & 0.0 & 0.0 & 0.0 & 0.0 & 0.0 & 15.4 & 19.2 & 23.6 & 28.2 & 31.0 & 8.4 & 7.2 & $\grbb{5.4}$ & $\grbb{5.2}$ & $\grbb{5.8}$ \\ 
\midrule
\multirow{4}{*}{\rotatebox[origin=c]{90}{DGP6}}&\multirow{4}{*}{\rotatebox[origin=c]{90}{GARCH}}& \multirow{4}{*}{\rotatebox[origin=c]{90}{SKST}}&2 & $\grbb{5.0}$ & $\grbb{4.8}$ & $\grbb{5.0}$ & $\grbb{3.6}$ & - & 2.4 & 1.0 & 0.8 & 0.6 & 1.8 & $\grbb{4.6}$ & $\grbb{5.6}$ & 7.6 & $\grbb{6.4}$ & 8.6 & $\grbb{3.4}$ & $\grbb{3.4}$ & $\grbb{4.6}$ & 1.4 & 2.6 \\ 
&&&10  & $\grbb{5.4}$ & $\grbb{3.4}$ & $\grbb{3.0}$ & $\grbb{3.4}$ & - & 0.0 & 0.0 & 0.0 & 0.0 & 0.0 & $\grbb{6.2}$ & $\grbb{6.6}$ & 8.2 & 7.6 & 8.8 & $\grbb{5.6}$ & $\grbb{3.2}$ & $\grbb{4.2}$ & $\grbb{3.4}$ & 2.6 \\ 
&&&50 & $\grbb{3.2}$ & $\grbb{4.6}$ & $\grbb{4.6}$ & $\grbb{3.6}$ & - & 0.0 & 0.0 & 0.0 & 0.0 & 0.0 & 10.2 & 10.2 & 11.2 & 12.2 & 13.6 & $\grbb{4.4}$ & $\grbb{5.0}$ & $\grbb{3.6}$ & 2.8 & $\grbb{4.2}$ \\ 
&&&100  & $\grbb{3.8}$ & $\grbb{4.4}$ & $\grbb{4.2}$ & $\grbb{4.2}$ & - & 0.0 & 0.0 & 0.0 & 0.0 & 0.0 & 15.0 & 20.8 & 25.0 & 28.0 & 33.8 & $\grbb{6.6}$ & $\grbb{6.4}$ & $\grbb{4.4}$ & $\grbb{6.8}$ & $\grbb{6.0}$ \\
\midrule\midrule
\multirow{4}{*}{\rotatebox[origin=c]{90}{DGP7}}&\multirow{4}{*}{\rotatebox[origin=c]{90}{AR}}& \multirow{4}{*}{\rotatebox[origin=c]{90}{N}}&2 & 13.0 & 21.6 & 49.4 & 64.8 & - & 7.4 & 7.4 & 8.8 & 9.2 & 15.2 & $\grbb{6.2}$ & $\grbb{5.8}$ & $\grbb{5.8}$ & $\grbb{4.8}$ & $\grbb{5.0}$ & $\grbb{5.0}$ & $\grbb{3.8}$ & $\grbb{3.2}$ & 2.4 & $\grbb{3.6}$ \\ 
&&&10  & 11.0 & 17.4 & 45.0 & 59.2 & - & 0.0 & 0.0 & 0.0 & 0.0 & 0.0 & $\grbb{4.4}$ & 2.8 & $\grbb{4.4}$ & $\grbb{6.0}$ & $\grbb{4.0}$ & $\grbb{4.2}$ & $\grbb{4.0}$ & $\grbb{4.0}$ & $\grbb{3.4}$ & $\grbb{4.0}$ \\ 
&&&50  & 8.0 & 15.6 & 36.0 & 42.0 & - & 0.0 & 0.0 & 0.0 & 0.0 & 0.0 & 8.8 & 12.0 & 14.6 & 14.2 & 16.2 & $\grbb{5.4}$ & $\grbb{3.2}$ & $\grbb{5.4}$ & $\grbb{4.8}$ & $\grbb{4.4}$ \\ 
&&&100  & 10.8 & 11.2 & 24.0 & 22.0 & - & 0.0 & 0.0 & 0.0 & 0.0 & 0.0 & 16.6 & 18.2 & 30.2 & 34.2 & 43.2 & $\grbb{7.0}$ & 7.6 & 8.2 & 8.8 & $\grbb{4.8}$ \\ 
\midrule
\multirow{4}{*}{\rotatebox[origin=c]{90}{DGP8}}&\multirow{4}{*}{\rotatebox[origin=c]{90}{AR}}& \multirow{4}{*}{\rotatebox[origin=c]{90}{ST}}&2 & 11.8 & 20.2 & 52.0 & 62.4 & - & $\grbb{5.6}$ & $\grbb{5.8}$ & 8.6 & 10.8 & 12.8 & $\grbb{5.6}$ & $\grbb{5.6}$ & 7.2 & 7.2 & 8.4 & $\grbb{5.0}$ & $\grbb{3.0}$ & $\grbb{4.4}$ & $\grbb{3.8}$ & 1.2 \\ 
&&&10 & 11.6 & 20.0 & 51.2 & 61.4 & - & 0.2 & 0.2 & 0.0 & 0.0 & 0.0 & $\grbb{6.2}$ & $\grbb{5.2}$ & $\grbb{6.8}$ & $\grbb{4.8}$ & 7.8 & $\grbb{4.6}$ & $\grbb{3.2}$ & $\grbb{4.6}$ & $\grbb{4.6}$ & $\grbb{3.0}$ \\ 
&&&50 & 10.8 & 17.2 & 35.8 & 44.4 & - & 0.0 & 0.0 & 0.0 & 0.0 & 0.0 &  9.6 & 10.6 & 13.0 & 14.2 & 14.4 & $\grbb{5.0}$ & $\grbb{4.6}$ & $\grbb{5.0}$ & $\grbb{4.2}$ & $\grbb{3.6}$ \\ 
&&&100 & 7.4 & 12.0 & 23.2 & 25.0 & - & 0.0 & 0.0 & 0.0 & 0.0 & 0.0 & 19.0 & 22.6 & 33.4 & 37.2 & 49.0 & $\grbb{5.4}$ & $\grbb{7.0}$ & 8.8 & $\grbb{6.2}$ & $\grbb{6.6}$ \\ 
\midrule
\multirow{4}{*}{\rotatebox[origin=c]{90}{DGP9}}&\multirow{4}{*}{\rotatebox[origin=c]{90}{AR}}& \multirow{4}{*}{\rotatebox[origin=c]{90}{SKST}}&2 & 13.8 & 26.0 & 52.6 & 66.2 & - & $\grbb{6.4}$ & $\grbb{6.6}$ & 8.8 & 11.4 & 10.8 & $\grbb{7.0}$ & $\grbb{6.6}$ & $\grbb{5.4}$ & $\grbb{5.2}$ & $\grbb{5.4}$ & $\grbb{5.4}$ & $\grbb{4.0}$ & 2.4 & 2.8 & $\grbb{3.8}$ \\ 
&&&10  & 8.4 & 17.2 & 48.4 & 64.2 & - & 0.2 & 0.0 & 0.0 & 0.0 & 0.0 & 7.2 & $\grbb{5.8}$ & $\grbb{5.8}$ & 7.2 & 8.4 & $\grbb{3.6}$ & $\grbb{3.0}$ & $\grbb{4.2}$ & 2.4 & $\grbb{3.6}$ \\ 
&&&50  & 9.2 & 18.8 & 39.0 & 43.8 & - & 0.0 & 0.0 & 0.0 & 0.0 & 0.0 & 9.2 & 11.4 & 13.6 & 12.8 & 17.6 & $\grbb{5.2}$ & $\grbb{5.4}$ & $\grbb{6.0}$ & $\grbb{4.0}$ & $\grbb{3.4}$ \\ 
&&&100  & 8.6 & 14.4 & 24.8 & 22.4 & - & 0.0 & 0.0 & 0.0 & 0.0 & 0.0 & 17.4 & 23.2 & 31.6 & 37.8 & 49.6 & 7.6 & 8.2 & 8.4 & 7.8 & $\grbb{6.6}$ \\
\midrule\midrule
\multirow{4}{*}{\rotatebox[origin=c]{90}{DGP10}}&\multirow{4}{*}{\rotatebox[origin=c]{90}{\small{AR-GARCH}}}& \multirow{4}{*}{\rotatebox[origin=c]{90}{N}}&2 & 10.6 & 21.6 & 49.6 & 66.0 & - & $\grbb{5.8}$ & $\grbb{7.0}$ & 8.2 & 10.8 & 13.4 & $\grbb{4.4}$ & $\grbb{5.2}$ & $\grbb{5.6}$ & $\grbb{6.0}$ & 7.2 & $\grbb{3.2}$ & $\grbb{3.4}$ & $\grbb{4.8}$ & 2.8 & $\grbb{3.2}$ \\ 
&&&10   & 10.8 & 18.8 & 48.2 & 62.0 & - & 0.2 & 0.0 & 0.2 & 0.0 & 0.2 & $\grbb{6.2}$ & $\grbb{5.6}$ & $\grbb{6.2}$ & 7.6 & $\grbb{6.2}$ & $\grbb{3.2}$ & 2.8 & 2.8 & $\grbb{4.2}$ & $\grbb{4.2}$ \\ 
&&&50   & 10.8 & 18.8 & 39.0 & 47.0 & - & 0.0 & 0.0 & 0.0 & 0.0 & 0.0 & 9.0 & 11.2 & 13.8 & 13.0 & 14.6 & $\grbb{3.6}$ & $\grbb{5.2}$ & $\grbb{5.8}$ & $\grbb{5.0}$ & $\grbb{5.4}$ \\ 
&&&100  & 8.0 & 10.8 & 26.6 & 24.4 & - & 0.0 & 0.0 & 0.0 & 0.0 & 0.0 & 15.8 & 20.8 & 33.2 & 35.2 & 50.0 & $\grbb{6.4}$ & $\grbb{6.8}$ & $\grbb{6.8}$ & $\grbb{5.0}$ & $\grbb{5.6}$ \\ 
\midrule
\multirow{4}{*}{\rotatebox[origin=c]{90}{DGP11}}&\multirow{4}{*}{\rotatebox[origin=c]{90}{\small{AR-GARCH}}}& \multirow{4}{*}{\rotatebox[origin=c]{90}{ST}}&2 & 10.8 & 23.2 & 49.0 & 67.8 & - & $\grbb{6.2}$ & 7.8 & 10.4 & 9.2 & 9.4 & 9.2 & 9.4 & $\grbb{7.0}$ & 8.6 & 7.2 & $\grbb{5.6}$ & $\grbb{3.4}$ & $\grbb{3.2}$ & $\grbb{3.2}$ & $\grbb{6.2}$ \\ 
&&&10 & 11.8 & 21.6 & 47.6 & 65.0 & - & 0.2 & 0.0 & 0.0 & 0.0 & 0.0 & 7.8 & 8.6 & 7.4 & 11.2 & 10.0 & $\grbb{5.0}$ & $\grbb{4.4}$ & 2.6 & $\grbb{3.2}$ & 2.2 \\ 
&&&50   & 9.0 & 17.2 & 42.2 & 45.8 & - & 0.0 & 0.0 & 0.0 & 0.0 & 0.0 & 8.6 & 13.4 & 14.8 & 18.2 & 19.8 & $\grbb{5.4}$ & $\grbb{5.0}$ & $\grbb{5.4}$ & $\grbb{6.6}$ & $\grbb{3.0}$ \\ 
&&&100  & 8.0 & 14.0 & 23.8 & 20.8 & - & 0.0 & 0.0 & 0.0 & 0.0 & 0.0 & 19.4 & 24.4 & 34.0 & 41.2 & 52.8 & 7.6 & $\grbb{6.6}$ & $\grbb{5.8}$ & $\grbb{5.4}$ & $\grbb{4.8}$ \\ 
\midrule
\multirow{4}{*}{\rotatebox[origin=c]{90}{DGP12}}&\multirow{4}{*}{\rotatebox[origin=c]{90}{\small{AR-GARCH}}}& \multirow{4}{*}{\rotatebox[origin=c]{90}{SKST}}&2 & 10.0 & 21.0 & 47.6 & 63.2 & - & $\grbb{5.0}$ & $\grbb{6.4}$ & 10.0 & 9.2 & 10.4 & $\grbb{4.6}$ & $\grbb{5.6}$ & 7.4 & 8.4 & 9.4 & 2.4 & $\grbb{3.4}$ & $\grbb{3.4}$ & 2.8 & $\grbb{4.2}$ \\ 
&&&10 & 11.4 & 22.0 & 47.8 & 60.8 & - & 0.0 & 0.0 & 0.0 & 0.0 & 0.2 & $\grbb{6.0}$ & $\grbb{6.0}$ & 7.4 & 8.4 &  9.8 & $\grbb{4.6}$ & $\grbb{3.8}$ & $\grbb{4.4}$ & $\grbb{4.0}$ & $\grbb{4.4}$ \\ 
&&&50 &  9.6 & 15.8 & 35.8 & 42.4 & - & 0.0 & 0.0 & 0.0 & 0.0 & 0.0 & 12.8 & 14.0 & 16.2 & 16.0 & 18.8 & $\grbb{3.8}$ & $\grbb{5.2}$ & $\grbb{7.0}$ & $\grbb{5.6}$ & $\grbb{6.2}$ \\ 
&&&100 & 7.8 & 16.2 & 20.0 & 21.8 & - & 0.0 & 0.0 & 0.0 & 0.0 & 0.0 & 18.4 & 26.2 & 38.8 & 42.8 & 52.6 & $\grbb{6.0}$ & $\grbb{5.2}$ & $\grbb{5.8}$ & $\grbb{6.0}$ & $\grbb{6.0}$ \\ 
\bottomrule
\end{tabular}}
\end{table}

\subsubsection{Size Results for $H_0^{\balpha}$}

Simulation results for the maximum Sharpe portfolio spanning $H_0^{\balpha}$  are reported in Table~\ref{tab:sizealpha}. 
While GRS, F1, BJ, and PY have an empirical size close to the nominal size of 5\% for the first six DGPs, they are all largely oversized in the presence of serial correlation, with empirical sizes sometimes approaching 80\%. Note also that the first three tests are not applicable when $N>T-K-1$, \eg, when $N=400$ and $T=250$. As for the joint test $H_0^{\balpha,\bdelta}$, GL is largely undersized for most configurations, even in the $i.i.d.$ Gaussian case.
The results of our proposed tests are satisfactory for all configurations and DGPs, especially BCS$^{\balpha}_2$.

\begin{table}[H]
\centering
\caption{\textbf{Size Results for $H_0^{\balpha}$}\\
The table reports the empirical size (over 500 replications) for eight tests
for the maximum Sharpe portfolio spanning under twelve DGPs and for four values of $K$ and $N$ (\ie, the number of benchmark and test assets, respectively). Values in bold are between 3\% and 7\%. A dash indicates that the test cannot be applied as $N>T-K-1$.}
\label{tab:sizealpha}
\scalebox{0.38}{
\begin{tabular}{p{0.1cm}p{0.1cm}llccccc|ccccc|ccccc|ccccc|ccccc|ccccc|ccccc}
\toprule
&&&&\multicolumn{5}{c}{GRS}
&\multicolumn{5}{c}{F1}
&\multicolumn{5}{c}{BJ}
&\multicolumn{5}{c}{PY}
&\multicolumn{5}{c}{GL}
&\multicolumn{5}{c}{BCS$^{\balpha,\bdelta}_0$}
&\multicolumn{5}{c}{BCS$^{\balpha,\bdelta}_2$}\\
\cmidrule(lr){5-9}
\cmidrule(lr){10-14}
\cmidrule(lr){15-19}
\cmidrule(lr){20-24}
\cmidrule(lr){25-29}
\cmidrule(lr){30-34}
\cmidrule(lr){35-39}
&&&$K\!\!\downarrow N\!\!\rightarrow$ 
& 2 & 10 & 50 & 100  & 400
& 2 & 10 & 50 & 100  & 400
& 2 & 10 & 50 & 100  & 400
& 2 & 10 & 50 & 100  & 400
& 2 & 10 & 50 & 100  & 400
& 2 & 10 & 50 & 100  & 400
& 2 & 10 & 50 & 100  & 400 \\
\midrule
\multirow{4}{*}{\rotatebox[origin=c]{90}{DGP1}}&\multirow{4}{*}{\rotatebox[origin=c]{90}{$i.i.d.$}}& \multirow{4}{*}{\rotatebox[origin=c]{90}{N}}&2 & $\grbb{5.2}$ & $\grbb{4.4}$ & $\grbb{5.4}$ & $\grbb{5.4}$ & - & $\grbb{5.2}$ & $\grbb{4.2}$ & $\grbb{5.2}$ & $\grbb{5.4}$ & - & $\grbb{5.0}$ & $\grbb{3.8}$ & $\grbb{4.8}$ & $\grbb{5.6}$ & - & $\grbb{5.8}$ & $\grbb{6.6}$ & $\grbb{6.2}$ & $\grbb{5.8}$ & $\grbb{6.0}$ & 1.0 & 0.2 & 0.0 & 0.0 & 0.2 & $\grbb{5.2}$ & $\grbb{4.8}$ & $\grbb{6.4}$ & $\grbb{6.6}$ & $\grbb{4.8}$ & 2.8 & $\grbb{4.0}$ & 2.4 & $\grbb{4.4}$ & $\grbb{4.0}$ \\ 
&&&10 & $\grbb{5.2}$ & $\grbb{4.2}$ & $\grbb{4.6}$ & $\grbb{5.0}$ & - & $\grbb{5.2}$ & $\grbb{4.2}$ & $\grbb{4.6}$ & $\grbb{5.0}$ & - & $\grbb{5.2}$ & $\grbb{4.4}$ & $\grbb{4.6}$ & $\grbb{5.0}$ & - & $\grbb{6.8}$ & $\grbb{7.0}$ & $\grbb{6.4}$ & $\grbb{5.4}$ & $\grbb{6.2}$ & 0.0 & 0.0 & 0.0 & 0.0 & 0.0 & $\grbb{5.2}$ & $\grbb{5.4}$ & 8.2 & $\grbb{7.0}$ & $\grbb{5.8}$ & $\grbb{4.0}$ & $\grbb{4.2}$ & $\grbb{4.2}$ & $\grbb{3.2}$ & 2.4 \\ 
&&&50  & $\grbb{5.8}$ & $\grbb{6.2}$ & $\grbb{5.4}$ & $\grbb{4.2}$ & - & $\grbb{5.8}$ & $\grbb{5.6}$ & $\grbb{5.0}$ & $\grbb{4.2}$ & - & $\grbb{6.0}$ & $\grbb{6.2}$ & $\grbb{5.0}$ & $\grbb{4.0}$ & - & $\grbb{6.4}$ & 7.4 & $\grbb{6.4}$ & $\grbb{4.2}$ & $\grbb{4.8}$ & 0.0 & 0.0 & 0.0 & 0.0 & 0.0 & 8.8 & 8.8 &  9.8 & 10.0 & 10.2 & $\grbb{4.6}$ & $\grbb{4.0}$ & $\grbb{4.6}$ & $\grbb{3.0}$ & $\grbb{4.0}$ \\ 
&&&100  & $\grbb{4.6}$ & $\grbb{4.0}$ & $\grbb{4.2}$ & $\grbb{4.2}$ & - & $\grbb{4.6}$ & $\grbb{4.0}$ & $\grbb{4.2}$ & $\grbb{4.2}$ & - & $\grbb{4.4}$ & $\grbb{4.4}$ & $\grbb{4.2}$ & $\grbb{4.4}$ & - & $\grbb{5.6}$ & 7.6 & $\grbb{6.2}$ & $\grbb{5.8}$ & $\grbb{3.8}$ & 0.0 & 0.0 & 0.0 & 0.0 & 0.0 & 8.8 & 14.4 & 20.0 & 19.0 & 21.8 & $\grbb{6.0}$ & $\grbb{5.8}$ & $\grbb{6.8}$ & 7.6 & $\grbb{4.4}$ \\ 
\midrule  
\multirow{4}{*}{\rotatebox[origin=c]{90}{DGP2}}&\multirow{4}{*}{\rotatebox[origin=c]{90}{$i.i.d.$}}& \multirow{4}{*}{\rotatebox[origin=c]{90}{ST}}&2 &$\grbb{4.2}$ & $\grbb{4.2}$ & $\grbb{5.8}$ & $\grbb{5.6}$ & - & $\grbb{4.2}$ & $\grbb{4.0}$ & $\grbb{5.4}$ & $\grbb{4.8}$ & - & $\grbb{4.4}$ & $\grbb{4.0}$ & $\grbb{5.8}$ & $\grbb{5.8}$ & - & $\grbb{6.8}$ & $\grbb{6.8}$ & $\grbb{7.0}$ & $\grbb{5.6}$ & $\grbb{6.2}$ & 0.8 & 0.4 & 0.6 & 0.2 & 0.2 & $\grbb{6.2}$ & 8.2 & 8.0 & $\grbb{6.0}$ & $\grbb{6.4}$ & $\grbb{4.0}$ & $\grbb{3.8}$ & 2.8 & $\grbb{3.0}$ & $\grbb{3.4}$ \\ 
&&&10 & $\grbb{4.6}$ & $\grbb{4.4}$ & $\grbb{5.6}$ & $\grbb{5.6}$ & - & $\grbb{4.4}$ & $\grbb{4.2}$ & $\grbb{5.6}$ & $\grbb{5.0}$ & - & $\grbb{4.2}$ & $\grbb{4.8}$ & $\grbb{6.2}$ & $\grbb{5.8}$ & - & $\grbb{6.8}$ & 7.6 & 8.4 & $\grbb{6.6}$ & $\grbb{7.0}$ & 0.0 & 0.0 & 0.0 & 0.0 & 0.0 & $\grbb{6.4}$ & $\grbb{6.0}$ & 7.2 & $\grbb{6.0}$ & $\grbb{5.2}$ & $\grbb{4.2}$ & $\grbb{4.4}$ & $\grbb{4.2}$ & $\grbb{4.6}$ & $\grbb{4.0}$ \\ 
&&&50  & $\grbb{3.2}$ & $\grbb{5.6}$ & $\grbb{4.4}$ & 7.4 & - & $\grbb{3.2}$ & $\grbb{5.6}$ & $\grbb{4.4}$ & 7.2 & - & $\grbb{3.4}$ & $\grbb{5.8}$ & $\grbb{5.0}$ & 7.4 & - & $\grbb{5.8}$ & $\grbb{7.0}$ & $\grbb{4.0}$ & $\grbb{5.6}$ & $\grbb{5.6}$ & 0.0 & 0.0 & 0.0 & 0.0 & 0.0 & $\grbb{7.0}$ & 9.4 & 9.0 &  9.8 &  9.6 & $\grbb{5.6}$ & $\grbb{4.6}$ & $\grbb{4.0}$ & $\grbb{3.8}$ & $\grbb{3.4}$ \\ 
&&&100 & $\grbb{3.4}$ & $\grbb{4.8}$ & $\grbb{5.6}$ & $\grbb{5.8}$ & - & $\grbb{3.4}$ & $\grbb{4.8}$ & $\grbb{5.4}$ & $\grbb{5.4}$ & - & $\grbb{3.6}$ & $\grbb{4.4}$ & $\grbb{6.0}$ & $\grbb{5.6}$ & - & $\grbb{6.2}$ & 7.4 & 9.0 & $\grbb{6.0}$ & $\grbb{5.6}$ & 0.0 & 0.0 & 0.0 & 0.0 & 0.0 & 13.4 & 17.8 & 20.8 & 19.0 & 22.4 & $\grbb{5.8}$ & $\grbb{5.8}$ & $\grbb{4.4}$ & $\grbb{6.2}$ & $\grbb{6.2}$ \\ 
\midrule  
\multirow{4}{*}{\rotatebox[origin=c]{90}{DGP3}}&\multirow{4}{*}{\rotatebox[origin=c]{90}{$i.i.d.$}}& \multirow{4}{*}{\rotatebox[origin=c]{90}{SKST}}&2 & $\grbb{4.6}$ & $\grbb{5.6}$ & $\grbb{4.6}$ & $\grbb{5.6}$ & - & $\grbb{4.6}$ & $\grbb{5.4}$ & $\grbb{4.6}$ & $\grbb{5.2}$ & - & $\grbb{4.4}$ & $\grbb{5.6}$ & $\grbb{4.4}$ & $\grbb{5.8}$ & - & $\grbb{7.0}$ & $\grbb{6.4}$ & $\grbb{3.8}$ & $\grbb{4.8}$ & $\grbb{4.6}$ & 0.2 & 0.0 & 0.2 & 0.2 & 0.4 & $\grbb{5.2}$ & $\grbb{5.4}$ & $\grbb{5.4}$ & $\grbb{5.4}$ & 7.6 & $\grbb{4.4}$ & $\grbb{4.4}$ & $\grbb{3.8}$ & $\grbb{4.6}$ & $\grbb{3.2}$ \\ 
&&&10& $\grbb{4.8}$ & $\grbb{5.8}$ & $\grbb{4.4}$ & $\grbb{5.4}$ & - & $\grbb{4.8}$ & $\grbb{5.8}$ & $\grbb{4.4}$ & $\grbb{5.4}$ & - & $\grbb{4.8}$ & $\grbb{6.2}$ & $\grbb{4.4}$ & $\grbb{5.4}$ & - & 7.6 & 7.4 & $\grbb{5.0}$ & $\grbb{5.2}$ & $\grbb{3.2}$ & 0.0 & 0.0 & 0.0 & 0.0 & 0.0 & $\grbb{6.4}$ & 7.8 & $\grbb{6.0}$ & $\grbb{3.8}$ & $\grbb{6.2}$ & $\grbb{6.2}$ & $\grbb{4.0}$ & $\grbb{3.4}$ & 2.4 & $\grbb{4.0}$ \\ 
&&&50& $\grbb{4.6}$ & $\grbb{5.4}$ & $\grbb{4.6}$ & $\grbb{6.2}$ & - & $\grbb{4.6}$ & $\grbb{5.4}$ & $\grbb{4.6}$ & $\grbb{6.2}$ & - & $\grbb{4.4}$ & $\grbb{5.8}$ & $\grbb{4.4}$ & $\grbb{6.2}$ & - & $\grbb{6.0}$ & 8.4 & $\grbb{6.4}$ & $\grbb{6.8}$ & $\grbb{5.4}$ & 0.0 & 0.0 & 0.0 & 0.0 & 0.0 & 7.2 & 10.0 & 11.2 & 12.0 & 9.4 & $\grbb{4.4}$ & $\grbb{3.2}$ & $\grbb{4.2}$ & $\grbb{5.4}$ & $\grbb{3.6}$ \\ 
&&&100 & $\grbb{4.4}$ & $\grbb{6.2}$ & $\grbb{4.4}$ & $\grbb{3.4}$ & - & $\grbb{4.4}$ & $\grbb{6.2}$ & $\grbb{4.4}$ & $\grbb{3.0}$ & - & $\grbb{3.8}$ & $\grbb{5.4}$ & $\grbb{5.0}$ & $\grbb{4.4}$ & - & $\grbb{6.0}$ & $\grbb{6.8}$ & $\grbb{6.0}$ & $\grbb{5.4}$ & $\grbb{5.4}$ & 0.0 & 0.0 & 0.0 & 0.0 & 0.0 & 10.0 & 15.4 & 19.8 & 20.8 & 27.4 & $\grbb{5.0}$ & $\grbb{3.6}$ & $\grbb{6.4}$ & $\grbb{5.4}$ & $\grbb{7.0}$ \\ 
\midrule  \midrule  
\multirow{4}{*}{\rotatebox[origin=c]{90}{DGP4}}&\multirow{4}{*}{\rotatebox[origin=c]{90}{GARCH}}& \multirow{4}{*}{\rotatebox[origin=c]{90}{N}}&2 & $\grbb{5.0}$ & $\grbb{5.6}$ & $\grbb{4.0}$ & $\grbb{4.0}$ & - & $\grbb{5.0}$ & $\grbb{5.2}$ & $\grbb{4.0}$ & $\grbb{4.0}$ & - & $\grbb{5.6}$ & $\grbb{5.4}$ & $\grbb{4.0}$ & $\grbb{4.0}$ & - & $\grbb{6.2}$ & $\grbb{5.4}$ & $\grbb{4.8}$ & $\grbb{4.0}$ & $\grbb{4.8}$ & 0.4 & 0.4 & 0.4 & 0.4 & 0.6 & $\grbb{3.6}$ & $\grbb{3.4}$ & $\grbb{4.8}$ & $\grbb{4.4}$ & $\grbb{5.0}$ & $\grbb{6.4}$ & $\grbb{3.0}$ & $\grbb{3.4}$ & $\grbb{3.2}$ & $\grbb{3.2}$ \\ 
&&&10  & $\grbb{5.2}$ & $\grbb{4.0}$ & $\grbb{5.0}$ & $\grbb{3.0}$ & - & $\grbb{5.2}$ & $\grbb{3.8}$ & $\grbb{5.0}$ & $\grbb{3.0}$ & - & $\grbb{5.0}$ & $\grbb{4.4}$ & $\grbb{4.6}$ & $\grbb{3.2}$ & - & $\grbb{6.0}$ & $\grbb{6.0}$ & $\grbb{6.4}$ & $\grbb{5.8}$ & $\grbb{4.8}$ & 0.0 & 0.0 & 0.0 & 0.0 & 0.0 & $\grbb{4.4}$ & $\grbb{5.8}$ & $\grbb{6.2}$ & $\grbb{4.6}$ & $\grbb{5.0}$ & $\grbb{5.2}$ & $\grbb{3.2}$ & $\grbb{3.6}$ & 2.4 & $\grbb{5.0}$ \\ 
&&&50 & $\grbb{5.2}$ & $\grbb{6.4}$ & $\grbb{3.8}$ & $\grbb{5.2}$ & - & $\grbb{4.8}$ & $\grbb{6.2}$ & $\grbb{3.8}$ & $\grbb{5.0}$ & - & $\grbb{5.0}$ & $\grbb{5.2}$ & $\grbb{4.2}$ & $\grbb{5.6}$ & - & $\grbb{7.0}$ & 8.0 & $\grbb{5.2}$ & $\grbb{7.0}$ & $\grbb{6.0}$ & 0.0 & 0.0 & 0.0 & 0.0 & 0.0 & 8.0 & 8.8 & $\grbb{7.0}$ & 8.0 & 8.2 & $\grbb{4.6}$ & $\grbb{4.2}$ & 2.6 & $\grbb{4.8}$ & $\grbb{3.2}$ \\ 
&&&100  & $\grbb{4.8}$ & $\grbb{4.6}$ & 8.6 & $\grbb{6.4}$ & - & $\grbb{4.8}$ & $\grbb{4.6}$ & 8.0 & $\grbb{6.4}$ & - & $\grbb{5.0}$ & $\grbb{4.8}$ & 8.2 & $\grbb{6.4}$ & - & 8.0 & $\grbb{6.8}$ & 8.4 & 8.0 & $\grbb{6.6}$ & 0.0 & 0.0 & 0.0 & 0.0 & 0.0 & 12.2 & 13.0 & 18.0 & 21.0 & 27.2 & 7.4 & $\grbb{5.8}$ & $\grbb{5.4}$ & $\grbb{6.6}$ & $\grbb{4.4}$ \\ 
 \midrule  
\multirow{4}{*}{\rotatebox[origin=c]{90}{DGP5}}&\multirow{4}{*}{\rotatebox[origin=c]{90}{GARCH}}& \multirow{4}{*}{\rotatebox[origin=c]{90}{ST}}&2 & $\grbb{5.0}$ & $\grbb{3.6}$ & $\grbb{6.0}$ & $\grbb{5.2}$ & - & $\grbb{5.0}$ & $\grbb{3.4}$ & $\grbb{5.8}$ & $\grbb{5.2}$ & - & $\grbb{4.8}$ & $\grbb{4.4}$ & $\grbb{5.6}$ & $\grbb{5.0}$ & - & 7.4 & $\grbb{7.0}$ & 8.0 & 7.4 & $\grbb{5.4}$ & 1.2 & 0.0 & 0.2 & 0.2 & 0.2 & 7.8 & $\grbb{5.4}$ & $\grbb{6.8}$ & $\grbb{6.0}$ & $\grbb{6.6}$ & $\grbb{4.4}$ & 2.6 & $\grbb{4.4}$ & $\grbb{3.0}$ & $\grbb{4.0}$ \\ 
&&&10  & $\grbb{4.2}$ & $\grbb{4.8}$ & $\grbb{5.2}$ & $\grbb{5.6}$ & - & $\grbb{4.2}$ & $\grbb{4.8}$ & $\grbb{4.8}$ & $\grbb{5.6}$ & - & $\grbb{4.6}$ & $\grbb{4.8}$ & $\grbb{4.8}$ & $\grbb{5.4}$ & - & $\grbb{6.0}$ & $\grbb{5.8}$ & $\grbb{7.0}$ & $\grbb{6.2}$ & $\grbb{5.2}$ & 0.0 & 0.0 & 0.0 & 0.0 & 0.0 & $\grbb{5.8}$ &  9.6 & $\grbb{6.6}$ & $\grbb{6.6}$ & $\grbb{6.0}$ & $\grbb{5.0}$ & $\grbb{3.8}$ & $\grbb{4.6}$ & $\grbb{4.6}$ & $\grbb{4.2}$ \\ 
&&&50 & $\grbb{3.4}$ & $\grbb{4.2}$ & $\grbb{5.8}$ & $\grbb{3.8}$ & - & $\grbb{3.2}$ & $\grbb{4.2}$ & $\grbb{5.6}$ & $\grbb{3.8}$ & - & $\grbb{3.6}$ & $\grbb{4.0}$ & $\grbb{5.4}$ & $\grbb{4.2}$ & - & $\grbb{6.6}$ & $\grbb{5.4}$ & $\grbb{6.8}$ & $\grbb{5.6}$ & $\grbb{5.4}$ & 0.0 & 0.0 & 0.0 & 0.0 & 0.0 & 8.2 &  9.8 & 12.6 & 10.8 & 11.4 & $\grbb{5.4}$ & 7.6 & $\grbb{3.6}$ & $\grbb{3.6}$ & $\grbb{4.0}$ \\ 
&&&100 & $\grbb{5.2}$ & $\grbb{5.6}$ & $\grbb{4.0}$ & $\grbb{3.8}$ & - & $\grbb{5.2}$ & $\grbb{5.6}$ & $\grbb{4.0}$ & $\grbb{3.8}$ & - & $\grbb{5.6}$ & $\grbb{5.4}$ & $\grbb{4.4}$ & $\grbb{4.0}$ & - & $\grbb{6.6}$ & $\grbb{5.8}$ & $\grbb{5.4}$ & $\grbb{5.2}$ & $\grbb{7.0}$ & 0.0 & 0.0 & 0.0 & 0.0 & 0.0 & 12.6 & 16.2 & 21.6 & 24.2 & 28.4 & $\grbb{3.8}$ & $\grbb{5.2}$ & $\grbb{5.8}$ & $\grbb{4.6}$ & $\grbb{6.0}$ \\ 
 \midrule  
\multirow{4}{*}{\rotatebox[origin=c]{90}{DGP6}}&\multirow{4}{*}{\rotatebox[origin=c]{90}{GARCH}}& \multirow{4}{*}{\rotatebox[origin=c]{90}{SKST}}&2 & $\grbb{4.0}$ & $\grbb{4.2}$ & $\grbb{5.6}$ & $\grbb{4.0}$ & - & $\grbb{4.0}$ & $\grbb{4.2}$ & $\grbb{5.6}$ & $\grbb{3.8}$ & - & $\grbb{4.2}$ & $\grbb{4.0}$ & $\grbb{5.4}$ & $\grbb{3.8}$ & - & 7.2 & 7.2 & 8.2 & $\grbb{5.4}$ & $\grbb{4.8}$ & 0.8 & 0.2 & 0.2 & 0.0 & 0.2 & $\grbb{5.8}$ & $\grbb{6.4}$ & $\grbb{6.0}$ & 7.4 & 10.2 & $\grbb{5.2}$ & $\grbb{4.2}$ & $\grbb{4.6}$ & $\grbb{3.2}$ & $\grbb{5.0}$ \\ 
&&&10 & $\grbb{6.0}$ & $\grbb{5.6}$ & $\grbb{5.4}$ & $\grbb{3.2}$ & - & $\grbb{6.0}$ & $\grbb{5.4}$ & $\grbb{5.4}$ & $\grbb{3.2}$ & - & $\grbb{6.2}$ & $\grbb{5.6}$ & $\grbb{5.4}$ & $\grbb{4.0}$ & - & $\grbb{5.8}$ & 8.2 & $\grbb{6.4}$ & $\grbb{5.0}$ & $\grbb{6.0}$ & 0.0 & 0.0 & 0.0 & 0.0 & 0.0 & $\grbb{6.4}$ & $\grbb{6.6}$ & 7.2 & $\grbb{6.4}$ & 8.4 & $\grbb{4.0}$ & $\grbb{3.0}$ & $\grbb{3.0}$ & $\grbb{4.8}$ & 2.8 \\ 
&&&50 & $\grbb{3.8}$ & $\grbb{4.6}$ & $\grbb{4.4}$ & 1.0 & - & $\grbb{3.8}$ & $\grbb{4.6}$ & $\grbb{4.2}$ & 0.8 & - & $\grbb{4.0}$ & $\grbb{4.8}$ & $\grbb{4.2}$ & 1.0 & - & $\grbb{6.6}$ & $\grbb{6.6}$ & $\grbb{4.0}$ & $\grbb{5.4}$ & $\grbb{6.2}$ & 0.0 & 0.0 & 0.0 & 0.0 & 0.0 & 8.0 & 10.2 & 10.2 & 10.0 & 11.8 & $\grbb{4.0}$ & $\grbb{4.2}$ & $\grbb{3.8}$ & $\grbb{4.2}$ & $\grbb{3.2}$ \\ 
&&&100 & $\grbb{5.0}$ & $\grbb{5.4}$ & $\grbb{3.2}$ & $\grbb{3.8}$ & - & $\grbb{5.0}$ & $\grbb{5.2}$ & $\grbb{3.2}$ & $\grbb{3.8}$ & - & $\grbb{4.6}$ & $\grbb{5.0}$ & $\grbb{3.6}$ & $\grbb{3.6}$ & - & $\grbb{5.2}$ & $\grbb{7.0}$ & $\grbb{5.4}$ & $\grbb{5.8}$ & $\grbb{6.0}$ & 0.0 & 0.0 & 0.0 & 0.0 & 0.0 & 11.0 & 16.8 & 20.8 & 22.8 & 29.6 & $\grbb{5.8}$ & $\grbb{4.6}$ & $\grbb{5.2}$ & $\grbb{4.8}$ & $\grbb{6.0}$ \\ 
 \midrule   \midrule  
\multirow{4}{*}{\rotatebox[origin=c]{90}{DGP7}}&\multirow{4}{*}{\rotatebox[origin=c]{90}{AR}}& \multirow{4}{*}{\rotatebox[origin=c]{90}{N}}&2 & 15.4 & 27.0 & 60.6 & 76.4 & - & 15.4 & 26.8 & 59.6 & 76.0 & - & 16.0 & 25.6 & 60.2 & 76.4 & - & 16.0 & 24.2 & 52.0 & 73.8 &  99.8 & $\grbb{3.8}$ & $\grbb{3.0}$ & $\grbb{4.8}$ & $\grbb{5.0}$ & $\grbb{6.8}$ & $\grbb{5.6}$ & $\grbb{5.4}$ & $\grbb{4.6}$ & $\grbb{5.0}$ & $\grbb{3.8}$ & $\grbb{3.8}$ & $\grbb{3.6}$ & $\grbb{4.4}$ & $\grbb{3.0}$ & $\grbb{3.8}$ \\ 
&&&10 & 13.8 & 24.0 & 57.2 & 70.8 & - & 13.8 & 24.0 & 57.0 & 70.2 & - & 14.0 & 24.2 & 57.4 & 70.6 & - & 13.0 & 20.8 & 50.4 & 72.8 &  99.6 & 0.0 & 0.0 & 0.0 & 0.0 & 0.0 & $\grbb{5.0}$ & $\grbb{5.2}$ & $\grbb{5.4}$ & 7.2 & $\grbb{6.2}$ & $\grbb{4.6}$ & $\grbb{3.2}$ & 2.6 & $\grbb{3.0}$ & $\grbb{3.2}$ \\ 
&&&50 & 11.2 & 19.6 & 43.2 & 54.6 & - & 11.2 & 19.2 & 43.0 & 53.6 & - & 11.8 & 19.4 & 44.0 & 55.0 & - & 14.0 & 18.6 & 41.8 & 60.6 & 96.6 & 0.0 & 0.0 & 0.0 & 0.0 & 0.0 & 9.2 & 12.6 & 12.4 & 13.4 & 17.4 & $\grbb{4.4}$ & $\grbb{5.8}$ & $\grbb{5.0}$ & $\grbb{5.2}$ & $\grbb{5.2}$ \\ 
&&&100 &  9.6 & 14.0 & 28.8 & 29.4 & - &  9.6 & 14.0 & 28.4 & 29.2 & - &  9.8 & 14.0 & 29.8 & 29.8 & - & 13.2 & 17.4 & 32.4 & 48.2 & 87.6 & 0.0 & 0.0 & 0.0 & 0.0 & 0.0 & 15.8 & 21.0 & 32.4 & 37.6 & 49.0 & $\grbb{6.0}$ & $\grbb{6.6}$ & 8.6 & 8.6 & $\grbb{6.6}$ \\ 
\midrule 
\multirow{4}{*}{\rotatebox[origin=c]{90}{DGP8}}&\multirow{4}{*}{\rotatebox[origin=c]{90}{AR}}& \multirow{4}{*}{\rotatebox[origin=c]{90}{ST}}&2 & 14.6 & 24.0 & 62.0 & 72.8 & - & 14.6 & 23.8 & 60.8 & 72.6 & - & 13.8 & 24.4 & 61.6 & 74.2 & - & 14.6 & 22.8 & 54.8 & 73.0 &  99.6 & $\grbb{3.2}$ & 2.2 & $\grbb{4.2}$ & $\grbb{4.8}$ & $\grbb{5.6}$ & $\grbb{5.8}$ & $\grbb{4.4}$ & $\grbb{6.4}$ & $\grbb{7.0}$ & $\grbb{6.4}$ & $\grbb{5.4}$ & $\grbb{3.0}$ & $\grbb{3.8}$ & 2.4 & $\grbb{3.0}$ \\ 
&&&10 & 14.4 & 26.4 & 60.8 & 71.2 & - & 14.2 & 26.2 & 60.4 & 71.0 & - & 14.8 & 26.2 & 62.0 & 72.4 & - & 16.8 & 26.0 & 55.8 & 71.6 & 99.4 & 0.2 & 0.0 & 0.0 & 0.0 & 0.0 & $\grbb{5.2}$ & $\grbb{5.2}$ & 7.2 & $\grbb{6.0}$ & $\grbb{6.2}$ & $\grbb{3.6}$ & $\grbb{4.2}$ & $\grbb{4.6}$ & $\grbb{3.0}$ & $\grbb{3.0}$ \\ 
&&&50& 11.4 & 18.2 & 42.4 & 51.8 & - & 11.4 & 18.0 & 41.0 & 51.6 & - & 11.4 & 18.4 & 42.6 & 53.6 & - & 12.0 & 17.8 & 39.0 & 58.8 & 97.8 & 0.0 & 0.0 & 0.0 & 0.0 & 0.0 & 10.2 & 13.2 & 12.8 & 13.0 & 14.0 & $\grbb{4.6}$ & $\grbb{5.0}$ & $\grbb{5.4}$ & $\grbb{5.2}$ & $\grbb{3.6}$ \\ 
&&&100 &  9.6 & 12.6 & 29.2 & 27.8 & - &  9.6 & 12.6 & 29.2 & 27.8 & - &  9.6 & 13.2 & 30.2 & 29.4 & - &  9.8 & 14.4 & 31.8 & 46.6 & 89.6 & 0.0 & 0.0 & 0.0 & 0.0 & 0.0 & 16.2 & 23.4 & 35.4 & 41.2 & 54.4 & 8.0 & $\grbb{5.4}$ & $\grbb{4.8}$ & $\grbb{5.6}$ & 8.4 \\ 
\midrule  
\multirow{4}{*}{\rotatebox[origin=c]{90}{DGP9}}&\multirow{4}{*}{\rotatebox[origin=c]{90}{AR}}& \multirow{4}{*}{\rotatebox[origin=c]{90}{SKST}}&2 & 15.6 & 28.4 & 62.2 & 75.4 & - & 15.6 & 28.2 & 61.2 & 75.0 & - & 15.2 & 27.4 & 61.4 & 75.8 & - & 15.8 & 25.8 & 58.6 & 76.4 & 99.4 & $\grbb{3.6}$ & $\grbb{3.0}$ & $\grbb{4.8}$ & $\grbb{5.2}$ & $\grbb{5.4}$ & $\grbb{6.8}$ & $\grbb{6.4}$ & $\grbb{6.2}$ & $\grbb{6.0}$ & $\grbb{5.0}$ & $\grbb{3.8}$ & $\grbb{4.8}$ & $\grbb{4.4}$ & $\grbb{3.8}$ & 1.4 \\ 
&&&10 & 12.2 & 22.8 & 55.6 & 72.0 & - & 12.2 & 22.6 & 55.2 & 71.8 & - & 12.2 & 23.2 & 57.2 & 72.0 & - & 14.2 & 20.8 & 52.8 & 73.0 & 99.4 & 0.2 & 0.0 & 0.0 & 0.0 & 0.0 & $\grbb{4.8}$ & $\grbb{5.6}$ & $\grbb{6.2}$ & $\grbb{6.4}$ & $\grbb{6.4}$ & $\grbb{4.8}$ & $\grbb{3.6}$ & $\grbb{4.6}$ & $\grbb{3.6}$ & 2.2 \\ 
&&&50  & 8.6 & 18.8 & 45.0 & 54.6 & - & 8.6 & 18.8 & 44.6 & 54.4 & - & 8.6 & 19.4 & 47.2 & 55.8 & - &  9.8 & 18.8 & 40.6 & 60.4 & 97.4 & 0.0 & 0.0 & 0.0 & 0.0 & 0.0 & 8.4 & 11.4 & 15.0 & 15.8 & 18.8 & $\grbb{5.8}$ & $\grbb{5.6}$ & $\grbb{3.0}$ & $\grbb{4.0}$ & $\grbb{4.0}$ \\ 
&&&100 & 10.2 & 16.8 & 32.8 & 29.0 & - & 10.2 & 16.8 & 32.6 & 29.0 & - &  9.8 & 16.8 & 33.0 & 28.6 & - & 11.4 & 16.4 & 30.0 & 47.4 & 90.8 & 0.0 & 0.0 & 0.0 & 0.0 & 0.0 & 17.8 & 22.4 & 30.4 & 40.6 & 51.6 & $\grbb{4.8}$ & 7.8 & $\grbb{7.0}$ & 8.0 & 7.6 \\
\midrule  \midrule
\multirow{4}{*}{\rotatebox[origin=c]{90}{DGP10}}&\multirow{4}{*}{\rotatebox[origin=c]{90}{\small{AR-GARCH}}}& \multirow{4}{*}{\rotatebox[origin=c]{90}{N}}&2 & 13.2 & 25.0 & 58.0 & 76.4 & - & 13.2 & 24.6 & 57.6 & 76.4 & - & 13.0 & 24.6 & 57.6 & 78.0 & - & 13.6 & 25.8 & 52.0 & 73.0 &  99.8 & 2.8 & 2.6 & 2.8 & $\grbb{5.0}$ & $\grbb{5.8}$ & $\grbb{4.4}$ & $\grbb{4.0}$ & $\grbb{5.0}$ & $\grbb{4.8}$ & $\grbb{5.4}$ & $\grbb{7.0}$ & $\grbb{3.2}$ & $\grbb{4.0}$ & $\grbb{3.2}$ & 2.4 \\ 
&&&10 & 10.0 & 23.4 & 57.4 & 75.6 & - &  9.8 & 23.2 & 56.8 & 73.8 & - & 10.6 & 23.4 & 56.6 & 74.0 & - & 13.4 & 23.6 & 52.0 & 72.2 &  99.6 & 0.2 & 0.0 & 0.0 & 0.0 & 0.0 & $\grbb{4.4}$ & $\grbb{5.0}$ & $\grbb{5.2}$ & $\grbb{4.8}$ & $\grbb{6.6}$ & $\grbb{4.4}$ & $\grbb{4.8}$ & $\grbb{3.4}$ & $\grbb{3.0}$ & $\grbb{4.2}$ \\ 
&&&50 & 11.8 & 25.4 & 49.6 & 58.0 & - & 11.8 & 25.2 & 48.8 & 57.2 & - & 11.8 & 26.2 & 49.8 & 59.4 & - & 14.0 & 22.8 & 44.0 & 63.0 & 97.0 & 0.0 & 0.0 & 0.0 & 0.0 & 0.0 & 8.4 & 11.8 & 12.6 & 13.2 & 14.2 & $\grbb{4.4}$ & $\grbb{4.4}$ & $\grbb{3.6}$ & $\grbb{4.2}$ & $\grbb{3.6}$ \\ 
&&&100 & 9.4 & 14.6 & 32.0 & 31.0 & - & 9.4 & 14.4 & 31.2 & 31.0 & - & 9.2 & 14.6 & 30.8 & 30.4 & - & 11.0 & 16.0 & 33.8 & 51.6 & 91.0 & 0.0 & 0.0 & 0.0 & 0.0 & 0.0 & 14.8 & 20.0 & 34.2 & 40.8 & 55.8 & 7.6 & $\grbb{7.0}$ & $\grbb{6.2}$ & 8.4 & $\grbb{5.6}$ \\ 
\midrule  
\multirow{4}{*}{\rotatebox[origin=c]{90}{DGP11}}&\multirow{4}{*}{\rotatebox[origin=c]{90}{\small{AR-GARCH}}}& \multirow{4}{*}{\rotatebox[origin=c]{90}{ST}}&2 & 14.0 & 24.8 & 60.6 & 75.8 & - & 14.0 & 24.2 & 59.4 & 75.0 & - & 14.2 & 25.4 & 61.0 & 76.8 & - & 15.4 & 22.0 & 51.8 & 75.0 &  99.6 & $\grbb{3.0}$ & 2.8 & $\grbb{6.2}$ & $\grbb{3.6}$ & $\grbb{4.6}$ & $\grbb{7.0}$ & $\grbb{5.6}$ & $\grbb{6.4}$ & $\grbb{6.2}$ & 7.8 & $\grbb{3.4}$ & $\grbb{3.2}$ & $\grbb{4.0}$ & $\grbb{3.8}$ & $\grbb{3.2}$ \\ 
&&&10 & 11.6 & 24.0 & 58.4 & 72.4 & - & 11.6 & 23.6 & 57.2 & 72.0 & - & 11.4 & 25.8 & 58.6 & 73.6 & - & 13.2 & 24.6 & 50.8 & 72.8 & 99.2 & 0.0 & 0.0 & 0.0 & 0.0 & 0.0 & $\grbb{6.6}$ & 11.0 & 9.2 & 7.2 & 7.6 & $\grbb{4.0}$ & 2.6 & $\grbb{4.6}$ & $\grbb{5.0}$ & $\grbb{3.2}$ \\ 
&&&50 & 10.6 & 20.2 & 41.8 & 55.6 & - & 10.6 & 19.6 & 41.4 & 55.2 & - & 11.4 & 19.6 & 42.6 & 56.2 & - & 13.4 & 21.2 & 42.0 & 60.0 & 98.0 & 0.0 & 0.0 & 0.0 & 0.0 & 0.0 &  9.8 & 12.6 & 17.6 & 16.0 & 20.0 & $\grbb{5.6}$ & 7.4 & $\grbb{4.6}$ & $\grbb{4.8}$ & $\grbb{5.4}$ \\ 
&&&100 & 10.8 & 17.4 & 28.4 & 25.4 & - & 10.8 & 17.2 & 28.2 & 25.2 & - & 11.4 & 16.8 & 30.4 & 27.2 & - & 12.4 & 16.8 & 31.0 & 48.0 & 89.4 & 0.0 & 0.0 & 0.0 & 0.0 & 0.0 & 16.6 & 23.0 & 35.2 & 39.6 & 54.2 & $\grbb{4.4}$ & $\grbb{5.0}$ & $\grbb{7.0}$ & $\grbb{5.6}$ & 7.6 \\ 
\midrule  
\multirow{4}{*}{\rotatebox[origin=c]{90}{DGP12}}&\multirow{4}{*}{\rotatebox[origin=c]{90}{\small{AR-GARCH}}}& \multirow{4}{*}{\rotatebox[origin=c]{90}{SKST}}&2 &  11.4 & 29.4 & 61.2 & 74.8 & - & 11.2 & 29.2 & 60.6 & 74.4 & - & 11.2 & 28.6 & 61.6 & 74.8 & - & 14.2 & 25.0 & 53.2 & 72.8 &  99.6 & 2.4 & $\grbb{3.0}$ & $\grbb{5.0}$ & $\grbb{4.8}$ & $\grbb{5.2}$ & $\grbb{5.4}$ & 7.2 & $\grbb{6.6}$ & 8.0 & 10.0 & $\grbb{4.4}$ & $\grbb{3.2}$ & 2.8 & $\grbb{3.8}$ & $\grbb{4.4}$ \\ 
&&&10 & 14.4 & 26.4 & 59.6 & 72.2 & - & 14.4 & 26.2 & 59.2 & 71.6 & - & 14.0 & 26.8 & 59.8 & 71.2 & - & 12.6 & 23.4 & 50.8 & 68.6 & 99.0 & 0.0 & 0.0 & 0.0 & 0.0 & 0.2 & 7.6 & $\grbb{6.6}$ & $\grbb{6.6}$ & $\grbb{7.0}$ & 9.0 & $\grbb{5.2}$ & $\grbb{5.2}$ & $\grbb{3.2}$ & $\grbb{4.8}$ & 2.6 \\ 
&&&50 & 12.0 & 21.4 & 41.4 & 52.8 & - & 12.0 & 21.2 & 41.0 & 52.0 & - & 12.2 & 20.2 & 41.8 & 53.2 & - & 14.2 & 20.0 & 38.6 & 59.6 & 97.2 & 0.0 & 0.0 & 0.0 & 0.0 & 0.0 & 10.0 & 14.6 & 15.4 & 17.2 & 19.4 & $\grbb{4.8}$ & $\grbb{3.8}$ & $\grbb{4.4}$ & $\grbb{6.4}$ & $\grbb{5.0}$ \\ 
&&&100 & 7.8 & 15.0 & 27.6 & 27.0 & - & 7.8 & 14.8 & 26.6 & 27.0 & - & 7.8 & 15.0 & 27.0 & 27.6 & - & 11.2 & 16.2 & 33.2 & 46.6 & 87.8 & 0.0 & 0.0 & 0.0 & 0.0 & 0.0 & 15.2 & 23.8 & 35.4 & 41.0 & 55.6 & 8.8 & $\grbb{5.8}$ & $\grbb{6.6}$ & 8.6 & 9.2 \\ 
\bottomrule
\end{tabular}}
\end{table}

\subsubsection{Size Results for $H_0^{\bdelta}$}

Results for the global minimum-variance portfolio hypothesis $H_0^{\bdelta}$ are reported in Table~\ref{tab:sizedelta}. 
While KM and F2 are not applicable for $N=400$, they have an empirical size close to the nominal size of 5\% for the four considered numbers of test assets $N$ when $K$ is small (\ie, $K=2$ or $10$). For $K=50$ and $100$, both tests are slightly oversized in the presence of serial correlation. Like for the other two simulations, our proposed BCS$^{\bdelta}_2$ test has, overall, the closest empirical size to the nominal size of 5\%, even when $N$ is much larger than the sample size.

\begin{table}[H]
\centering
\caption{\textbf{Size Results for $H_0^{\bdelta}$}\\
The table reports the empirical size (over 500 replications) for five tests
for the global minimum-variance spanning under twelve DGPs and four values of $K$ and $N$ (\ie, the number of benchmark and test assets, respectively). Values in bold are between 3\% and 7\%. A dash indicates that the test cannot be applied as $N>T-K-1$.}
\label{tab:sizedelta}
\singlespacing
\scalebox{0.6}{
\begin{tabular}{p{0.1cm}p{0.1cm}llccccc|ccccc|ccccc|ccccc|ccccc}
\toprule
&&&&\multicolumn{5}{c}{KM}
&\multicolumn{5}{c}{F2}
&\multicolumn{5}{c}{BCS$^{\bdelta}_0$}
&\multicolumn{5}{c}{BCS$^{\bdelta}_2$}\\
\cmidrule(lr){5-9}
\cmidrule(lr){10-14}
\cmidrule(lr){15-19}
\cmidrule(lr){20-24}
&&&$K\!\!\downarrow N\!\!\rightarrow$ 
& 2 & 10 & 50 & 100 & 400 
& 2 & 10 & 50 & 100 & 400 
& 2 & 10 & 50 & 100 & 400 
& 2 & 10 & 50 & 100 & 400 \\
\midrule
\multirow{4}{*}{\rotatebox[origin=c]{90}{DGP1}}&\multirow{4}{*}{\rotatebox[origin=c]{90}{$i.i.d.$}}& \multirow{4}{*}{\rotatebox[origin=c]{90}{N}}&2 & $\grbb{5.4}$ & $\grbb{5.4}$ & $\grbb{5.6}$ & $\grbb{6.2}$ & - & $\grbb{4.8}$ & $\grbb{5.6}$ & $\grbb{5.2}$ & $\grbb{6.2}$ & - & $\grbb{5.4}$ & $\grbb{7.0}$ & $\grbb{6.2}$ & $\grbb{6.4}$ & $\grbb{5.2}$ & $\grbb{4.6}$ & $\grbb{4.0}$ & $\grbb{3.8}$ & $\grbb{4.4}$ & $\grbb{3.4}$ \\ 
&&&10& $\grbb{5.6}$ & $\grbb{4.8}$ & $\grbb{5.4}$ & $\grbb{6.0}$ & - & $\grbb{5.6}$ & $\grbb{4.8}$ & $\grbb{5.8}$ & $\grbb{5.2}$ & - & $\grbb{5.2}$ & $\grbb{5.0}$ & $\grbb{5.2}$ & $\grbb{5.4}$ & $\grbb{6.0}$ & $\grbb{5.0}$ & 2.4 & $\grbb{4.4}$ & $\grbb{4.4}$ & $\grbb{3.2}$ \\ 
&&&50& $\grbb{5.8}$ & $\grbb{5.2}$ & $\grbb{6.0}$ & $\grbb{6.0}$ & - & $\grbb{5.8}$ & $\grbb{5.4}$ & $\grbb{5.4}$ & $\grbb{5.6}$ & - & $\grbb{5.6}$ & 8.4 & 9.0 &  9.6 & 10.0 & $\grbb{4.0}$ & $\grbb{6.4}$ & 2.8 & 2.8 & $\grbb{4.8}$ \\ 
&&&100& $\grbb{4.0}$ & $\grbb{4.6}$ & $\grbb{5.0}$ & $\grbb{6.2}$ & - & $\grbb{4.4}$ & $\grbb{4.4}$ & $\grbb{4.6}$ & $\grbb{6.6}$ & - & 7.6 & 13.6 & 15.2 & 20.2 & 28.6 & $\grbb{4.0}$ & $\grbb{6.0}$ & $\grbb{4.0}$ & $\grbb{7.0}$ & $\grbb{6.6}$ \\ 
\midrule  
\multirow{4}{*}{\rotatebox[origin=c]{90}{DGP2}}&\multirow{4}{*}{\rotatebox[origin=c]{90}{$i.i.d.$}}& \multirow{4}{*}{\rotatebox[origin=c]{90}{ST}}&2 & $\grbb{4.6}$ & $\grbb{5.6}$ & $\grbb{5.8}$ & $\grbb{5.0}$ & - & $\grbb{4.4}$ & $\grbb{5.2}$ & $\grbb{5.8}$ & $\grbb{5.6}$ & - & $\grbb{3.0}$ & $\grbb{3.8}$ & $\grbb{6.6}$ & $\grbb{6.2}$ & $\grbb{5.0}$ & $\grbb{3.6}$ & 2.8 & $\grbb{4.2}$ & $\grbb{4.6}$ & $\grbb{3.2}$ \\ 
&&&10 & $\grbb{5.0}$ & $\grbb{4.4}$ & $\grbb{5.2}$ & $\grbb{4.2}$ & - & $\grbb{4.8}$ & $\grbb{4.0}$ & $\grbb{4.4}$ & $\grbb{4.4}$ & - & $\grbb{4.2}$ & $\grbb{5.0}$ & $\grbb{6.6}$ & 7.4 & $\grbb{6.8}$ & $\grbb{4.0}$ & $\grbb{5.0}$ & $\grbb{4.0}$ & $\grbb{3.4}$ & 2.8 \\ 
&&&50& $\grbb{5.4}$ & 7.2 & $\grbb{5.0}$ & $\grbb{5.8}$ & - & $\grbb{5.4}$ & $\grbb{6.8}$ & $\grbb{5.0}$ & $\grbb{5.6}$ & - & 7.4 &  9.6 & 11.6 & 11.2 & 13.2 & $\grbb{4.4}$ & $\grbb{3.0}$ & $\grbb{3.4}$ & $\grbb{4.0}$ & $\grbb{5.2}$ \\ 
&&&100 & $\grbb{4.2}$ & $\grbb{6.0}$ & $\grbb{5.6}$ & $\grbb{6.4}$ & - & $\grbb{4.4}$ & $\grbb{6.4}$ & $\grbb{5.2}$ & $\grbb{6.2}$ & - & 11.2 & 14.8 & 15.8 & 17.0 & 24.8 & 7.2 & $\grbb{6.2}$ & $\grbb{6.6}$ & $\grbb{6.8}$ & $\grbb{6.2}$ \\ 
\midrule  
\multirow{4}{*}{\rotatebox[origin=c]{90}{DGP2}}&\multirow{4}{*}{\rotatebox[origin=c]{90}{$i.i.d.$}}& \multirow{4}{*}{\rotatebox[origin=c]{90}{SKST}}&2 & $\grbb{6.0}$ & $\grbb{4.4}$ & $\grbb{4.0}$ & $\grbb{6.0}$ & - & $\grbb{6.2}$ & $\grbb{4.2}$ & $\grbb{4.2}$ & $\grbb{5.4}$ & - & $\grbb{5.8}$ & 7.2 & $\grbb{4.0}$ & $\grbb{5.8}$ & $\grbb{6.0}$ & $\grbb{4.4}$ & $\grbb{4.6}$ & $\grbb{4.4}$ & $\grbb{3.8}$ & $\grbb{4.2}$ \\ 
&&&10 & $\grbb{5.0}$ & $\grbb{5.0}$ & $\grbb{5.2}$ & $\grbb{4.4}$ & - & $\grbb{5.2}$ & $\grbb{4.8}$ & $\grbb{5.4}$ & $\grbb{4.8}$ & - & $\grbb{5.8}$ & $\grbb{6.4}$ & $\grbb{6.0}$ & $\grbb{6.4}$ & $\grbb{6.4}$ & $\grbb{4.4}$ & $\grbb{5.8}$ & $\grbb{4.0}$ & 2.8 & $\grbb{3.2}$ \\ 
&&&50& 7.2 & $\grbb{6.8}$ & $\grbb{5.2}$ & $\grbb{4.4}$ & - & $\grbb{6.6}$ & $\grbb{6.2}$ & $\grbb{5.0}$ & $\grbb{4.8}$ & - & 8.2 & 10.4 & 12.6 & 14.2 & 13.2 & $\grbb{6.6}$ & $\grbb{6.0}$ & $\grbb{4.4}$ & $\grbb{3.6}$ & $\grbb{4.6}$ \\ 
&&&100 & $\grbb{5.4}$ & $\grbb{4.0}$ & $\grbb{3.8}$ & $\grbb{6.2}$ & - & $\grbb{4.8}$ & $\grbb{4.0}$ & $\grbb{3.8}$ & $\grbb{6.4}$ & - & 12.0 & 18.0 & 20.2 & 21.6 & 26.2 & $\grbb{5.6}$ & $\grbb{5.8}$ & $\grbb{5.8}$ & $\grbb{5.0}$ & 7.8 \\ 
\midrule  \midrule
\multirow{4}{*}{\rotatebox[origin=c]{90}{DGP4}}&\multirow{4}{*}{\rotatebox[origin=c]{90}{GARCH}}& \multirow{4}{*}{\rotatebox[origin=c]{90}{N}}&2 & $\grbb{4.8}$ & $\grbb{5.4}$ & $\grbb{4.6}$ & $\grbb{5.4}$ & - & $\grbb{4.8}$ & $\grbb{5.6}$ & $\grbb{4.6}$ & $\grbb{5.2}$ & - & $\grbb{6.8}$ & $\grbb{6.0}$ & $\grbb{5.6}$ & $\grbb{5.2}$ & $\grbb{5.0}$ & $\grbb{3.8}$ & 2.8 & 1.8 & $\grbb{4.0}$ & $\grbb{4.4}$ \\ 
&&&10 & $\grbb{6.2}$ & $\grbb{3.2}$ & $\grbb{4.0}$ & $\grbb{4.2}$ & - & $\grbb{5.8}$ & $\grbb{3.6}$ & $\grbb{3.6}$ & $\grbb{4.6}$ & - & $\grbb{6.4}$ & $\grbb{4.4}$ & $\grbb{6.4}$ & $\grbb{5.0}$ & $\grbb{6.2}$ & $\grbb{3.6}$ & $\grbb{3.2}$ & $\grbb{3.4}$ & $\grbb{3.4}$ & $\grbb{3.6}$ \\ 
&&&50  & $\grbb{5.2}$ & $\grbb{4.0}$ & $\grbb{5.6}$ & $\grbb{6.0}$ & - & $\grbb{5.0}$ & $\grbb{3.8}$ & $\grbb{5.8}$ & $\grbb{5.6}$ & - & $\grbb{5.6}$ & 8.2 & 9.4 & 9.2 & 9.4 & $\grbb{4.0}$ & $\grbb{4.0}$ & $\grbb{4.0}$ & $\grbb{5.6}$ & 2.8 \\ 
&&&100  & $\grbb{4.6}$ & $\grbb{5.8}$ & $\grbb{5.8}$ & $\grbb{6.0}$ & - & $\grbb{4.4}$ & $\grbb{5.6}$ & $\grbb{5.4}$ & $\grbb{5.8}$ & - & 12.0 & 16.8 & 22.0 & 22.8 & 31.0 & $\grbb{6.0}$ & $\grbb{5.8}$ & $\grbb{5.2}$ & $\grbb{5.4}$ & $\grbb{5.8}$ \\ 
\midrule    
\multirow{4}{*}{\rotatebox[origin=c]{90}{DGP5}}&\multirow{4}{*}{\rotatebox[origin=c]{90}{GARCH}}& \multirow{4}{*}{\rotatebox[origin=c]{90}{ST}}&2 & $\grbb{4.0}$ & $\grbb{4.6}$ & $\grbb{5.0}$ & $\grbb{6.4}$ & - & $\grbb{4.4}$ & $\grbb{4.4}$ & $\grbb{4.4}$ & $\grbb{6.4}$ & - & 7.4 & 8.8 & 9.4 & 10.2 & 10.2 & $\grbb{3.8}$ & $\grbb{4.4}$ & $\grbb{3.6}$ & $\grbb{4.2}$ & $\grbb{3.8}$ \\ 
&&&10  & $\grbb{5.4}$ & $\grbb{3.0}$ & $\grbb{3.6}$ & $\grbb{6.2}$ & - & $\grbb{5.6}$ & 2.8 & $\grbb{3.8}$ & $\grbb{6.2}$ & - & $\grbb{6.6}$ & $\grbb{5.8}$ & 8.4 & 9.2 & 8.4 & $\grbb{6.0}$ & $\grbb{3.0}$ & $\grbb{4.0}$ & $\grbb{3.8}$ & $\grbb{3.0}$ \\ 
&&&50  & $\grbb{5.8}$ & $\grbb{6.2}$ & $\grbb{6.2}$ & 8.4 & - & $\grbb{6.0}$ & $\grbb{6.4}$ & $\grbb{6.8}$ & 7.8 & - & 7.8 & 12.4 & 12.2 & 13.8 & 12.6 & $\grbb{4.8}$ & $\grbb{5.6}$ & $\grbb{5.0}$ & $\grbb{3.2}$ & $\grbb{4.8}$ \\ 
&&&100  & $\grbb{3.8}$ & $\grbb{4.2}$ & $\grbb{5.6}$ & $\grbb{6.0}$ & - & $\grbb{4.0}$ & $\grbb{4.0}$ & $\grbb{6.0}$ & $\grbb{6.4}$ & - & 15.0 & 17.8 & 19.2 & 24.2 & 29.4 & $\grbb{6.4}$ & $\grbb{6.6}$ & $\grbb{6.4}$ & $\grbb{5.8}$ & $\grbb{6.2}$ \\ 
\midrule  
\multirow{4}{*}{\rotatebox[origin=c]{90}{DGP6}}&\multirow{4}{*}{\rotatebox[origin=c]{90}{GARCH}}& \multirow{4}{*}{\rotatebox[origin=c]{90}{SKST}}&2 & $\grbb{3.8}$ & $\grbb{5.0}$ & $\grbb{3.2}$ & $\grbb{4.6}$ & - & $\grbb{4.0}$ & $\grbb{4.6}$ & 2.8 & $\grbb{4.2}$ & - & $\grbb{4.4}$ & $\grbb{6.0}$ & 7.8 & $\grbb{6.8}$ & 8.0 & $\grbb{4.6}$ & $\grbb{5.0}$ & 2.8 & $\grbb{3.2}$ & 2.8 \\ 
&&&10 & $\grbb{4.6}$ & $\grbb{4.6}$ & $\grbb{3.6}$ & $\grbb{5.2}$ & - & $\grbb{5.4}$ & $\grbb{4.4}$ & $\grbb{3.8}$ & $\grbb{4.2}$ & - & $\grbb{6.8}$ & 8.2 & 8.6 & 8.0 & 11.2 & $\grbb{5.0}$ & $\grbb{5.0}$ & $\grbb{6.6}$ & $\grbb{4.6}$ & $\grbb{3.4}$ \\ 
&&&50 & $\grbb{4.6}$ & $\grbb{5.2}$ & $\grbb{6.2}$ & 7.4 & - & $\grbb{4.2}$ & $\grbb{4.8}$ & $\grbb{6.4}$ & $\grbb{7.0}$ & - & 10.0 & 9.4 & 10.8 & 13.8 & 12.8 & $\grbb{5.4}$ & $\grbb{5.0}$ & $\grbb{4.8}$ & $\grbb{4.6}$ & $\grbb{4.4}$ \\ 
&&&100 & $\grbb{4.4}$ & $\grbb{5.4}$ & $\grbb{5.4}$ & $\grbb{3.8}$ & - & $\grbb{4.8}$ & $\grbb{5.0}$ & $\grbb{6.2}$ & $\grbb{5.0}$ & - & 14.4 & 17.6 & 22.8 & 25.4 & 29.6 & $\grbb{6.4}$ & $\grbb{4.0}$ & $\grbb{5.8}$ & $\grbb{5.6}$ & $\grbb{5.8}$ \\ 
\midrule    \midrule
\multirow{4}{*}{\rotatebox[origin=c]{90}{DGP7}}&\multirow{4}{*}{\rotatebox[origin=c]{90}{AR}}& \multirow{4}{*}{\rotatebox[origin=c]{90}{N}}&2 & $\grbb{5.0}$ & 8.8 & 8.4 & 11.8 & - & $\grbb{6.0}$ & 9.0 & 9.0 & 11.0 & - & $\grbb{6.4}$ & $\grbb{6.2}$ & $\grbb{5.4}$ & $\grbb{6.2}$ & $\grbb{5.4}$ & $\grbb{5.6}$ & $\grbb{3.4}$ & $\grbb{5.0}$ & $\grbb{3.0}$ & $\grbb{4.8}$ \\ 
&&&10  & $\grbb{4.6}$ & $\grbb{5.0}$ & 8.4 & 10.0 & - & $\grbb{4.6}$ & $\grbb{4.4}$ & 8.2 &  9.8 & - & $\grbb{4.8}$ & $\grbb{3.0}$ & $\grbb{4.8}$ & $\grbb{5.2}$ & $\grbb{4.6}$ & $\grbb{4.6}$ & $\grbb{3.2}$ & 2.8 & $\grbb{4.6}$ & $\grbb{4.0}$ \\ 
&&&50  & $\grbb{4.8}$ & 7.4 & 9.4 & 8.0 & - & $\grbb{4.8}$ & 7.8 & 9.2 & 8.6 & - & 7.6 & 11.2 & 12.0 & 12.2 & 12.0 & $\grbb{3.8}$ & $\grbb{5.8}$ & $\grbb{3.6}$ & $\grbb{5.2}$ & $\grbb{5.4}$ \\ 
&&&100  & $\grbb{6.4}$ & $\grbb{5.8}$ & $\grbb{6.4}$ & 7.2 & - & $\grbb{6.6}$ & $\grbb{5.8}$ & $\grbb{6.4}$ & $\grbb{6.2}$ & - & 14.0 & 12.2 & 22.2 & 24.0 & 28.4 & $\grbb{5.4}$ & $\grbb{4.6}$ & $\grbb{6.6}$ & $\grbb{5.6}$ & $\grbb{5.4}$ \\ 
\midrule    
\multirow{4}{*}{\rotatebox[origin=c]{90}{DGP8}}&\multirow{4}{*}{\rotatebox[origin=c]{90}{AR}}& \multirow{4}{*}{\rotatebox[origin=c]{90}{ST}}&2 & $\grbb{6.0}$ & 7.4 & 7.8 & 12.4 & - & $\grbb{6.0}$ & $\grbb{6.6}$ & 7.4 & 11.6 & - & $\grbb{6.2}$ & $\grbb{6.8}$ & $\grbb{6.0}$ & $\grbb{6.6}$ & $\grbb{5.8}$ & $\grbb{5.2}$ & $\grbb{3.2}$ & $\grbb{4.2}$ & $\grbb{4.0}$ & $\grbb{4.0}$ \\ 
&&&10  & $\grbb{5.6}$ & $\grbb{6.0}$ & 9.0 & 11.8 & - & $\grbb{5.2}$ & $\grbb{5.2}$ & 9.4 & 11.4 & - & $\grbb{5.2}$ & $\grbb{5.8}$ & $\grbb{7.0}$ & $\grbb{5.0}$ & 8.2 & $\grbb{4.0}$ & $\grbb{4.2}$ & 2.6 & $\grbb{3.8}$ & $\grbb{3.8}$ \\ 
&&&50  & $\grbb{6.4}$ & 7.4 & 10.8 & 10.0 & - & $\grbb{6.4}$ & 7.2 & 11.4 & 9.4 & - & 8.0 & 9.0 & 11.0 & 13.0 & 11.4 & $\grbb{4.2}$ & $\grbb{4.6}$ & $\grbb{4.4}$ & $\grbb{4.8}$ & $\grbb{4.4}$ \\ 
&&&100  & $\grbb{4.8}$ & $\grbb{7.0}$ & 8.8 & $\grbb{7.0}$ & - & $\grbb{4.8}$ & $\grbb{6.4}$ & 8.4 & $\grbb{6.8}$ & - & 13.0 & 15.4 & 20.0 & 24.2 & 26.8 & $\grbb{5.2}$ & $\grbb{4.4}$ & $\grbb{6.0}$ & $\grbb{5.8}$ & $\grbb{5.8}$ \\ 
\midrule    
\multirow{4}{*}{\rotatebox[origin=c]{90}{DGP9}}&\multirow{4}{*}{\rotatebox[origin=c]{90}{AR}}& \multirow{4}{*}{\rotatebox[origin=c]{90}{SKST}}&2 & $\grbb{6.2}$ & 8.2 &  9.8 & 12.0 & - & $\grbb{6.0}$ & 7.6 & 11.4 & 12.0 & - & $\grbb{5.2}$ & $\grbb{7.0}$ & $\grbb{5.2}$ & $\grbb{5.4}$ & $\grbb{4.8}$ & $\grbb{4.6}$ & $\grbb{3.6}$ & $\grbb{3.2}$ & $\grbb{5.2}$ & $\grbb{3.0}$ \\ 
&&&10  & $\grbb{6.6}$ & 7.2 & 12.4 & 13.2 & - & $\grbb{6.4}$ & $\grbb{7.0}$ & 11.6 & 13.0 & - & 8.0 & $\grbb{5.8}$ & $\grbb{6.6}$ & $\grbb{7.0}$ & 9.2 & $\grbb{5.0}$ & $\grbb{6.4}$ & $\grbb{4.4}$ & $\grbb{3.6}$ & $\grbb{3.0}$ \\ 
&&&50  & 8.0 & 8.2 & 11.2 & 11.0 & - & 8.2 & 7.8 & 11.8 & 10.2 & - & 7.2 & 8.4 & 7.8 & 8.8 & 11.8 & $\grbb{4.2}$ & $\grbb{5.0}$ & $\grbb{3.0}$ & $\grbb{4.8}$ & $\grbb{3.2}$ \\ 
&&&100  & $\grbb{6.2}$ & $\grbb{6.0}$ & $\grbb{6.6}$ & 8.0 & - & $\grbb{5.8}$ & $\grbb{5.6}$ & $\grbb{7.0}$ & 8.2 & - & 13.0 & 16.8 & 23.2 & 28.2 & 32.2 & 7.2 & $\grbb{7.0}$ & $\grbb{5.6}$ & $\grbb{7.0}$ & $\grbb{4.4}$ \\
\midrule   \midrule
\multirow{4}{*}{\rotatebox[origin=c]{90}{DGP10}}&\multirow{4}{*}{\rotatebox[origin=c]{90}{\small{AR-GARCH}}}& \multirow{4}{*}{\rotatebox[origin=c]{90}{N}}&2 & $\grbb{5.8}$ & 8.0 & 10.2 & 11.2 & - & $\grbb{5.8}$ & 8.2 & 11.2 & 11.8 & - & $\grbb{6.2}$ & $\grbb{6.2}$ & $\grbb{6.6}$ & $\grbb{6.8}$ & 7.6 & $\grbb{4.4}$ & $\grbb{4.4}$ & 2.6 & $\grbb{3.8}$ & $\grbb{3.6}$ \\ 
&&&10 & $\grbb{6.4}$ & $\grbb{7.0}$ & 9.2 & 10.6 & - & $\grbb{5.8}$ & $\grbb{6.6}$ & 8.8 & 11.2 & - & 7.8 & $\grbb{6.0}$ & 8.0 & $\grbb{7.0}$ & 7.8 & $\grbb{5.2}$ & $\grbb{3.8}$ & $\grbb{4.8}$ & 2.8 & $\grbb{4.0}$ \\ 
&&&50& $\grbb{6.0}$ & $\grbb{5.4}$ & 10.0 & 9.2 & - & $\grbb{6.0}$ & $\grbb{5.4}$ & 10.0 & 9.4 & - & 7.8 & 10.2 & 14.4 & 13.0 & 12.8 & $\grbb{3.8}$ & $\grbb{5.2}$ & 2.8 & $\grbb{4.0}$ & $\grbb{3.2}$ \\ 
&&&100 & $\grbb{4.8}$ & $\grbb{6.6}$ & 7.6 & 8.2 & - & $\grbb{5.2}$ & $\grbb{6.6}$ & $\grbb{6.8}$ & 7.8 & - & 13.2 & 17.6 & 24.4 & 23.6 & 34.0 & $\grbb{5.2}$ & $\grbb{5.8}$ & $\grbb{4.8}$ & $\grbb{4.6}$ & $\grbb{4.8}$ \\ 
\midrule    
\multirow{4}{*}{\rotatebox[origin=c]{90}{DGP11}}&\multirow{4}{*}{\rotatebox[origin=c]{90}{\small{AR-GARCH}}}& \multirow{4}{*}{\rotatebox[origin=c]{90}{ST}}&2 & $\grbb{4.8}$ & 7.6 & 8.6 & 13.6 & - & $\grbb{4.8}$ & 8.2 & 8.8 & 13.8 & - & $\grbb{7.0}$ &  9.8 & 8.4 & 9.2 &  9.6 & $\grbb{3.2}$ & $\grbb{6.0}$ & $\grbb{3.2}$ & 1.8 & $\grbb{3.4}$ \\ 
&&&10  & 7.4 & $\grbb{6.2}$ & 8.0 & 12.4 & - & $\grbb{7.0}$ & $\grbb{6.6}$ & 8.0 & 13.0 & - & $\grbb{6.6}$ & $\grbb{7.0}$ &  9.6 & 10.8 & 11.4 & $\grbb{5.4}$ & $\grbb{4.0}$ & $\grbb{3.0}$ & $\grbb{3.8}$ & 2.6 \\ 
&&&50  & $\grbb{6.4}$ & 7.2 & 12.4 & 12.6 & - & $\grbb{6.4}$ & 7.8 & 12.2 & 11.8 & - & 9.4 & 12.0 & 14.6 & 16.0 & 16.0 & $\grbb{4.8}$ & $\grbb{4.6}$ & $\grbb{4.2}$ & $\grbb{3.0}$ & $\grbb{5.2}$ \\ 
&&&100  & $\grbb{5.2}$ & $\grbb{5.4}$ & 8.8 & 8.6 & - & $\grbb{5.4}$ & $\grbb{6.0}$ & 9.2 & 9.2 & - & 13.6 & 18.8 & 27.0 & 32.0 & 38.2 & 7.2 & $\grbb{6.6}$ & $\grbb{6.8}$ & 7.4 & $\grbb{5.2}$ \\ 
\midrule    
\multirow{4}{*}{\rotatebox[origin=c]{90}{DGP12}}&\multirow{4}{*}{\rotatebox[origin=c]{90}{\small{AR-GARCH}}}& \multirow{4}{*}{\rotatebox[origin=c]{90}{SKST}}&2 & $\grbb{6.4}$ & 7.6 & 10.0 & 12.8 & - & $\grbb{6.8}$ & 7.8 & 10.8 & 13.2 & - & $\grbb{4.2}$ & $\grbb{5.6}$ & $\grbb{6.4}$ & $\grbb{6.4}$ & 8.0 & $\grbb{3.4}$ & $\grbb{5.0}$ & $\grbb{3.6}$ & $\grbb{4.2}$ & $\grbb{3.0}$ \\ 
&&&10 & $\grbb{5.6}$ & $\grbb{6.4}$ & 8.6 & 11.2 & - & $\grbb{4.8}$ & $\grbb{6.6}$ & 8.6 & 10.6 & - & $\grbb{5.4}$ & 8.0 & 8.0 & 8.4 & 10.2 & $\grbb{4.4}$ & $\grbb{6.8}$ & $\grbb{4.2}$ & $\grbb{5.4}$ & $\grbb{5.2}$ \\ 
&&&50  & $\grbb{4.8}$ & 7.6 & 12.8 & 11.2 & - & $\grbb{5.2}$ & $\grbb{7.0}$ & 12.8 & 11.2 & - & 10.6 & 10.4 & 13.0 & 15.4 & 16.4 & $\grbb{5.8}$ & $\grbb{5.2}$ & $\grbb{4.2}$ & $\grbb{4.4}$ & $\grbb{5.8}$ \\ 
&&&100 & $\grbb{4.6}$ & $\grbb{6.0}$ & 7.4 & $\grbb{6.4}$ & - & $\grbb{4.8}$ & $\grbb{6.2}$ & 7.4 & $\grbb{7.0}$ & - & 15.2 & 18.6 & 27.2 & 30.0 & 35.2 & $\grbb{4.8}$ & $\grbb{4.4}$ & $\grbb{5.0}$ & 7.4 & $\grbb{5.0}$ \\ 
\bottomrule
\end{tabular}}
\end{table}

\subsection{Power Results}\label{sec:power}

In this section, we study the power of the tests considered in the previous section. To simulate data under the alternative, we consider a sparse setting where we set $\alpha_{i}=\delta_i=a$ for $i=1,\ldots,[N/2]$ with $a \in [-0.4,0.4]$ and $a=0$ for $i>[N/2]$, where $[x]$ denotes the integer part of $x$.

For the three tested null hypotheses, we focus on the most complete DGP, namely DGP12, for which returns have serial correlation, GARCH effects, and innovations follow an asymmetric Student distribution. 

\subsubsection{Power Results for $H_0^{\balpha,\bdelta}$}\label{sec:powerjoint}

\begin{figure}[H]
\centering
\caption{\textbf{Empirical Power of the MVS Tests}\\
The plots display the empirical power (over 500 replications) for the MVS tests $H_0^{\balpha,\bdelta}$ under the AR-GARCH(1,1) model with skewed Student-$t$ errors for 
various $K$ benchmark and $N$ test assets. We consider $K\in\{2,10,100\}$ and $N\in\{2,50,400\}$. For $N=400$, the test HK is not feasible. The solid horizontal lines correspond to the nominal level of 5\%.}
\includegraphics[width=1\linewidth]{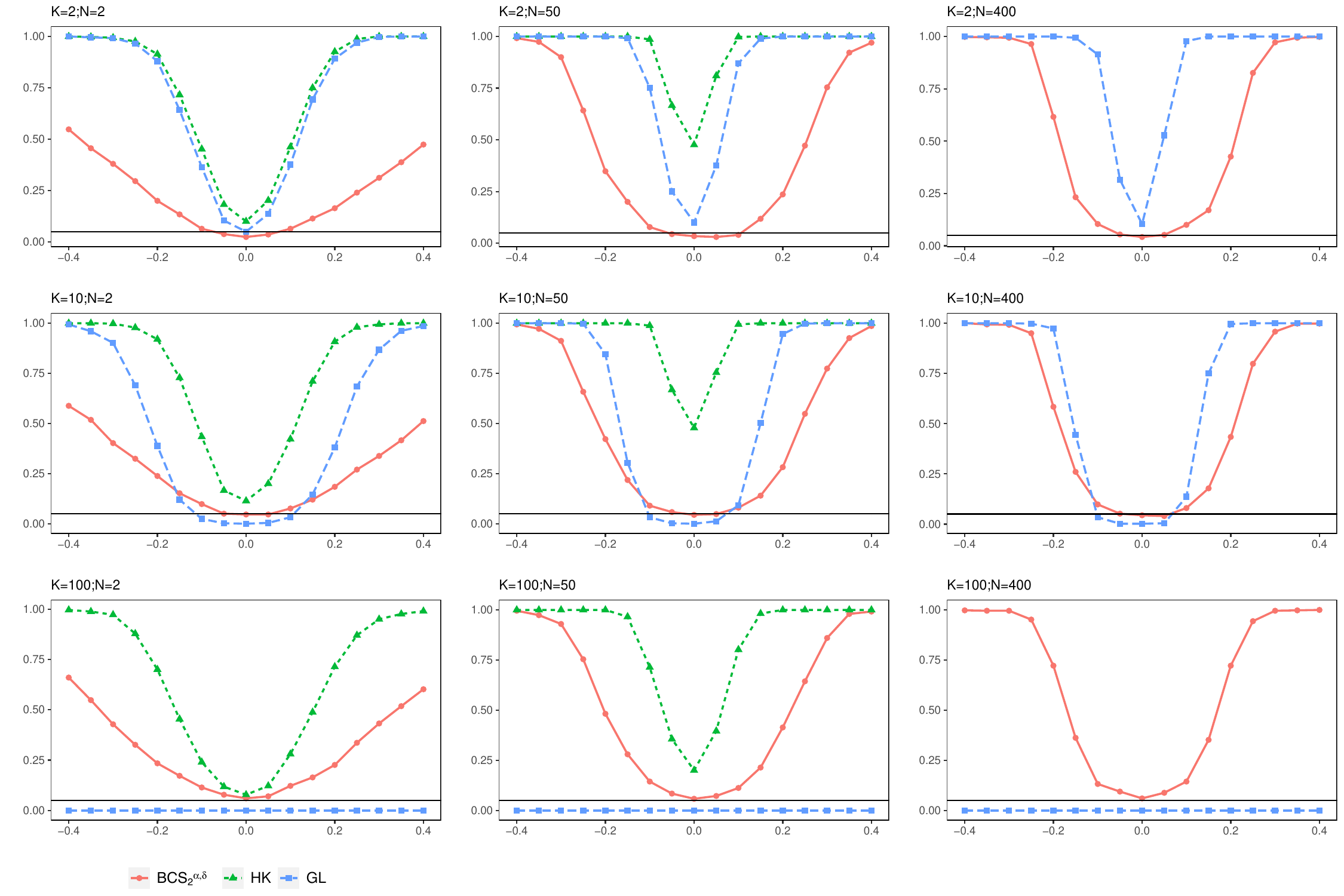}\\
\label{fig:powermvs}
\end{figure}

The empirical power functions for $H_0^{\balpha,\bdelta}$ (\ie, HK, GL, and BCS$^{\balpha,\bdelta}_2$) are plotted in Figure~\ref{fig:powermvs}. It is important to note that the rejection frequencies are not adjusted for the possible size distortion.
Recall that we have seen in Table~\ref{tab:sizeH0alphadelta} that HK is largely oversized in this case while GL is largely undersized except for $K=2$.

This figure demonstrates the contribution of the randomly-weighted batch-mean CCT with $L=2$. In the unlikely case $K=N=2$, this test is dominated by both HK and GL, although HK was found to be oversized in this case. HK displays the highest rejection frequencies, but those rejections are mostly due to the large size distortion of this test when $K$ and $N$ are large. While GL has slightly more power than BCS$^{\balpha,\bdelta}_2$ when $K$ is small, it has no power when $K$ is large (\ie, $K=100$). Importantly, BCS$^{\balpha,\bdelta}_2$ has similar power empirical curves, whatever the number of benchmark assets, for a given number of test assets and is found to have decent power in all configurations.     

\subsubsection{Power Results for $H_0^{\balpha}$}

\begin{figure}[H]
\centering
\caption{\textbf{Empirical Power of the Maximum Sharpe Portfolio Spanning Tests}\\
The plots display the empirical power (over 500 replications) of the maximum Sharpe portfolio spanning tests $H_0^{\balpha}$ under the AR-GARCH(1,1) model with skewed Student-$t$ errors for  $K$ benchmark and $N$ test assets. We consider $K\in\{2,10,100\}$ and $N\in\{2,50,400\}$. For $N=400$, the GRS, F1, and BJ tests are not feasible. The solid horizontal lines correspond to the nominal level of 5\%.}
\includegraphics[width=1\linewidth]{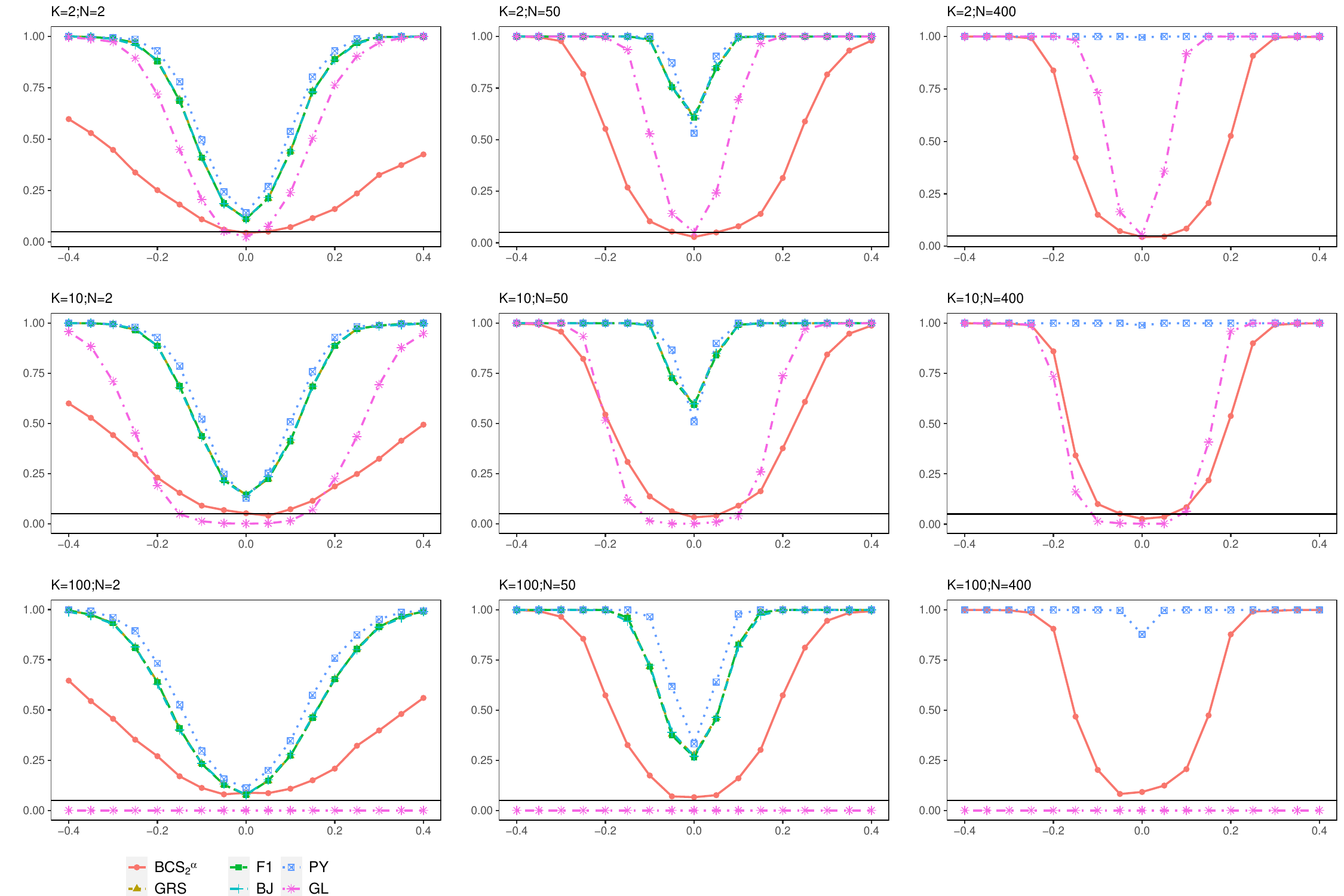}\\
\label{fig:poweralpha}
\end{figure}

The power curves of the maximum Sharpe portfolio spanning tests $H_0^{\balpha}$ are plotted in Figure~\ref{fig:poweralpha}. As for the joint test, the contribution of BCS$^{\balpha}_2$ is visible when the number of test assets is large. Indeed, unlike the GRS, F1, and BJ, this test is still applicable when $N>T$ and in this case, it has a decent size (unlike PY) and it has power against the null hypothesis $H_0^{\balpha}$ (unlike GL).

\subsubsection{Power Results for $H_0^{\bdelta}$}

\begin{figure}[H]
\centering
\caption{\textbf{Empirical Power of the Global Minimum-Variance Portfolio Spanning Tests}\\
The plots display the empirical power (over 500 replications) of the global minimum-variance portfolio spanning tests $H_0^{\bdelta}$ under the AR-GARCH(1,1) model with skewed Student-$t$ errors for various $K$ benchmark and $N$ test assets. We consider $K\in\{2,10,100\}$ and $N\in\{2,50,400\}$. For $N=400$, the KM and F2 tests are not feasible. The solid horizontal lines correspond to the nominal level of 5\%.}
\includegraphics[width=1\linewidth]{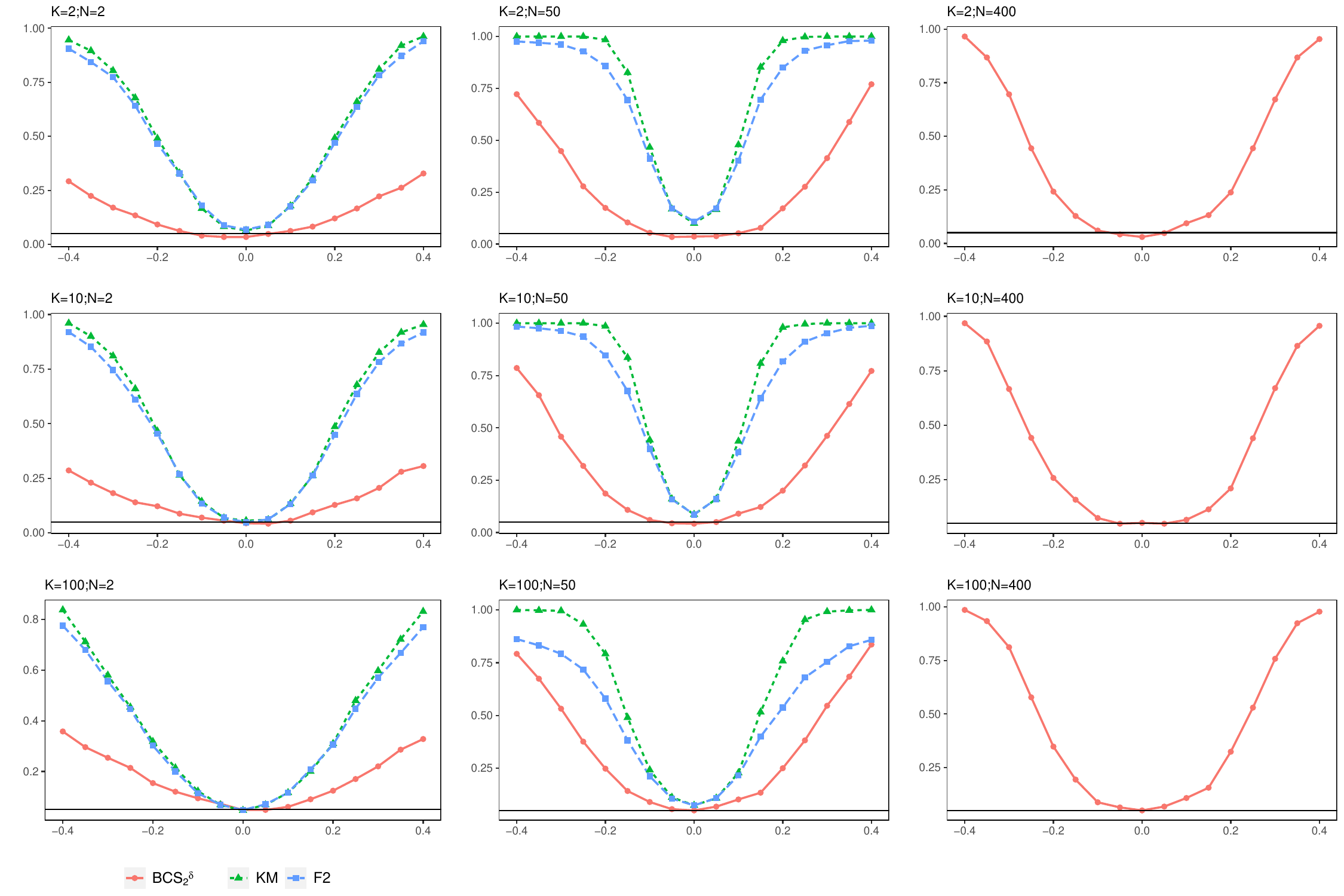}\\
\label{fig:powerdelta}
\end{figure}

Finally, the power curves of the minimum-variance portfolio spanning tests $H_0^{\bdelta}$ are plotted in Figure~\ref{fig:powerdelta}. When $N=2$, KM and F2 have much higher power than BCS$^{\bdelta}_2$. When $N$ is moderately large (\ie, $N=50$), rejection frequencies of KM and F2 are higher than BCS$^{\bdelta}_2$ but these two tests are slightly oversized unlike BCS$^{\bdelta}_2$. When $N=400$, the only applicable test is BCS$^{\bdelta}_2$, and the power curves are again insensitive to the number of benchmark assets.   

To sum up, the BCS$^{\balpha, \bdelta}_2$, BCS$^{\balpha}_2$ and BCS$^{\bdelta}_2$ tests are the ones retaining good properties in the presence of serial correlation and GARCH effects, even when the number of test and benchmark assets is large.

\section{Empirical Application}
\label{sec:empirics}

This section tests to which extent domestic equity investors could benefit from international diversification. Our empirical illustration uses daily returns from 2007 to 2022 on blue-chip stocks traded in North America and Europe. Specifically, at the begining of each year, we use stock returns of the historical constituents in the following equity indices: (i) S\&P 100 (U.S.), (ii) Euronext 100 (Eurozone), and (iii) SMI (Switzerland). We apply the three spanning 
tests $H_0^{\balpha,\bdelta}$, $H_0^{\balpha}$, and $H_0^{\bdelta}$ every year  to determine whether it is relevant to include assets from the other two countries. The number of benchmark and test assets varies slightly over time but remains a large-dimensional problem, and the sample size for each year is slightly less than 250 days; see the online appendix for details. 

For BCS$^{\blambda}_L$, we set $L=2$ as the empirical size was found to be satisfactory in the Monte Carlo study for $T=250$. We consider a fixed-width rule $\zeta = 1/3$, but results are robust to alternative choices. We compare our results for $H_0^{\balpha,\bdelta}$ and $H_0^{\balpha}$ with the GL approach using the default implementation with 500 boostrap replications. Results are reported in Table~\ref{tab:empirical}, where tests with a p-value lower that 5\% are highlighted by a symbol $\checkmark$ while inconclusive test outcomes (of the GL tests) are highlighted by a question mark'.

First, we find that the test $H_0^{\balpha,\bdelta}$ is rejected on average one-third 
of the time for each country and the rejections do not align among countries. Second, we see that the rejection of the MVS hypothesis comes from the potential of variance reduction in the domestic global minimum-variance portfolio. Rejections of $H_0^{\bdelta}$ drives the rejection of $H_0^{\balpha,\bdelta}$. On the contrary, the maximum Sharpe portfolio spanning test $H_0^{\balpha}$ is only rejected a few times. 

On the other hand, we are unable to consistently reject the null hypothesis $H_0^{\balpha,\bdelta}$ using GL, leading to inconclusive results when $K$ is moderately large (see columns S\&P 100 and Euronext 100). In slightly more than 25\% of the cases, GL yields inconclusive results when test assets are SMI constituents. However, we do not reject $H_0^{\balpha,\bdelta}$ using GL in 2007 and 2017 when test assets are S\&P 100 constituents. We reject $H_0^{\balpha,\bdelta}$ around 75\% of the cases when SMI constituents are the test assets. This behavior of the GL test statistics is expected as it becomes less informative as $K$ increases (see Section \ref{sec:powerjoint}). When focusing on $H_0^{\balpha}$, we are unable to reject the null hypothesis of Sharpe ratio portfolio spanning with any of the test assets universes under consideration using GL. Overall, we can conclude that the few rejections of $H_0^{\balpha,\bdelta}$ are driven by the rejection of $H_0^{\bdelta}$. However, unlike BCS$^{\blambda}_2$, we obtain fewer rejections of $H_0^{\balpha,\bdelta}$, which can be explained by the superior performance of BCS$^{\blambda}_2$ over GL in terms of size and power (see Sections \ref{sec:size} and \ref{sec:power}).

\begin{table}[H]
\centering
\caption{\textbf{Results of the MVS Tests Over Time for the Various Investment Universes}\\
This table reports the rejections at the 5\% significance level (highlighted by a symbol $\checkmark$) of the various MVS tests applied at the asset level. $H_0^{\balpha,\bdelta}$ is the MVS test, $H_0^{\balpha}$ is the maximum Sharpe ratio spanning test, and $H_0^{\bdelta}$ is the global minimum-variance spanning test. GL is the test by \citet{gungor2016multivariate} implemented with 500 bootstrap replications and BCS$^{\blambda}_2$ is the our BCS test with $L=2$. A question mark indicates an inconclusive test outcome.}
\label{tab:empirical}
\singlespacing
\scalebox{0.8}{
\begin{tabular}{lccccccccccccccc}
\toprule
&\multicolumn{5}{c}{S\&P 100}&\multicolumn{5}{c}{Euronext 100}&\multicolumn{5}{c}{SMI}\\
\cmidrule(lr){2-6}\cmidrule(lr){7-11}\cmidrule(lr){12-16}
&\multicolumn{2}{c}{GL}&\multicolumn{3}{c}{BCS$^{\blambda}_2$}
&\multicolumn{2}{c}{GL}&\multicolumn{3}{c}{BCS$^{\blambda}_2$}
&\multicolumn{2}{c}{GL}&\multicolumn{3}{c}{BCS$^{\blambda}_2$}\\
\cmidrule(lr){2-3}\cmidrule(lr){4-6}
\cmidrule(lr){7-8}\cmidrule(lr){9-11}
\cmidrule(lr){12-13}\cmidrule(lr){14-16}
Year 
& $H_0^{\balpha,\bdelta}$ & $H_0^{\balpha}$ &  $H_0^{\balpha,\bdelta}$ & $H_0^{\balpha}$ & $H_0^{\delta}$ 
& $H_0^{\balpha,\bdelta}$ & $H_0^{\balpha}$ &  $H_0^{\balpha,\bdelta}$ & $H_0^{\balpha}$ & $H_0^{\bdelta}$  
& $H_0^{\balpha,\bdelta}$ & $H_0^{\balpha}$ &  $H_0^{\balpha,\bdelta}$ & $H_0^{\balpha}$ & $H_0^{\bdelta}$ \\
\midrule
2007 &  &  &  &  &  & ? &  &  &  &\checkmark&\checkmark&  &\checkmark&  &  \\ 
2008 & ? &  &\checkmark&  &\checkmark& ? &  &\checkmark&  &  & ? &  &  &  &  \\ 
2009 & ? &  &\checkmark&\checkmark&\checkmark& ? &  &  &  &  & ? &  &  &  &  \\ 
2010 & ? &  &\checkmark&  &\checkmark& ? &  &  &  &  &\checkmark&  &\checkmark&  &\checkmark\\ 
2011 & ? &  &  &  &  & ? &  &\checkmark&  &\checkmark&\checkmark&  &\checkmark&  &\checkmark\\ 
2012 & ? &  &  &  &  & ? &  &  &  &  &\checkmark&  &  &  &\checkmark\\ 
2013 & ? &  &  &  &  & ? &  &\checkmark&  &  &\checkmark&  &\checkmark&  &\checkmark\\ 
2014 & ? &  &  &  &\checkmark& ? &  &\checkmark&  &\checkmark&\checkmark&  &  &\checkmark&  \\ 
2015 & ? &  &\checkmark&  &  & ? &  &  &  &\checkmark&\checkmark&  &\checkmark&  &\checkmark\\ 
2016 & ? &  &  &  &\checkmark& ? &  &\checkmark&  &\checkmark&\checkmark&  &\checkmark&  &  \\ 
2017 &  &  &  &  &\checkmark& ? & ? &  &  &\checkmark&\checkmark&  &  &  &\checkmark\\ 
2018 & ? &  &\checkmark&  &\checkmark& ? &  &  &  &\checkmark&\checkmark&  &\checkmark&  &\checkmark\\ 
2019 & ? &  &\checkmark&  &\checkmark& ? &  &\checkmark&  &  &\checkmark&  &  &  &\checkmark\\ 
2020 & ? &  &\checkmark&  &\checkmark& ? &  &  &  &  & ? &  &  &  &  \\ 
2021 & ? &  &  &  &\checkmark& ? &  &  &\checkmark&  &\checkmark&  &\checkmark&  &\checkmark\\ 
2022 & ? &  &\checkmark&  &\checkmark& ? &  &  &  &  & ? &  &  &  &  \\
\bottomrule
\end{tabular}}
\end{table}

\section{Conclusion}
\label{sec:conclusion}

The paper proposes a new framework for mean-variance spanning (MVS) testing that is more general and computationally efficient than existing methods. The proposed framework can be applied to any test-asset dimension and only requires stationary asset returns. It uses new moment conditions for spanning and tests them in two steps. First, it tests each component of the moment vector using robust Student-t tests based on the batch-mean method. Second, it combines the individual p-values using the Cauchy combination test (CCT) of \citet{liu2020cauchy}, which accounts for the cross-sectional dependence between the test statistics. Monte Carlo simulations show the proposed MVS tests have correct sizes and high power in most setups. Unlike state-of-the-art methods, they also work well on skewed, heteroscedastic, and fat-tailed data. The methodology is applied to test if combining blue-chip stocks traded in the U.S., Europe, and Switzerland can improve each country's domestic mean-variance efficient frontier. We find that the benefits of international diversification depend on economic conditions and vary across countries. We also highlight that the rejection of the MVS hypothesis originates from the potential to reduce variance within the domestic global minimum-variance portfolio. 


%

\section*{Acknowledgments}

David is grateful to IVADO and the Natural Sciences and Engineering Research Council of Canada (grant
RGPIN-2022-03767). Sébastien acknowledges the research support
of the French National Research Agency Grants ANR-17-EURE-0020 and ANR-21-CE26-0007-01. Rosnel acknowledges the financial support of Fin-ML. We thank Marie-Claude Beaulieu and Olivier Scaillet for their comments, and Richard Luger for providing us with the code for the GL test.

\newpage
\singlespacing

\newpage
\begin{titlepage}
\begin{center}
\vspace*{1cm}
\huge{
Online Appendix\\
High-Dimensional Mean-Variance Spanning Tests}
\vfill
\end{center}
\end{titlepage}

\pagenumbering{arabic} 
\setlength\parindent{0pt}
\setcounter{equation}{0}
\renewcommand*{\theequation}{A\arabic{equation}} 
\setcounter{table}{0}
\renewcommand*{\thetable}{A.\arabic{table}} 
\setcounter{figure}{0}
\renewcommand*{\thefigure}{A.\arabic{figure}} 
\setcounter{section}{0}
\renewcommand*{\thesection}{\Roman{section}} 
\setcounter{subsection}{0}
\renewcommand*{\thesubsection}{\thesection.\Alph{subsection}} 
\setcounter{page}{1}
\renewcommand*{\thepage}{\Roman{page}} 
\doublespacing

\section{Additional Lemma and Proofs}

Lemma \ref{lemma1} derives the decomposition of a precision matrix using nodewize regressions. Similar results can also be found in \citet{peng2009partial}.

\begin{lemma}\label{lemma1}
Consider the random vector $\bx_t\in \mathbb{R}^d$ and the following sequence of $d$ regressions $x_{j,t}=\sum_{i=1,i\neq j}^{d}\theta_{i,j}x_{i,t}+v_{j,t}$ with $\mathbb{E}[\bx_{-j,t}v_{j,t}]=0$ and $\mathbb{E}[v_{j,t}]=\bO_{(d-1)\times 1}$  for $j=1,\ldots,d$. Let $\mathbb{V}[v_{i,t}]=g^2_i$ for $i=1,\ldots,d$,  $\bG\equiv\operatorname{Diag}(g^2_1,\ldots,g^2_d)$ and $\bTheta\equiv(\theta_{i,j})$. 
\begin{enumerate}
\item  Using the exogeneity conditions $\mathbb{E}[\bx_{-j,t}v_{j,t}]=\bO_{(d-1)\times 1}$ for any $j$, and $k\neq j$, one can deduce that $\mathbb{E}[v_k,v_j]=\mathbb{E}[v_k(x_{j,t}-\sum_{i=1,i\neq j}^{d}\theta_{i,j}x_{i,t})]=-\theta_{j,k}\mathbb{E}[v_k,x_k]=-\theta_{j,k}\mathbb{E}[v_k,v_k]=-\theta_{j,k}\mathbb{V}(v_k)=-\theta_{j,k}g_k$.	
\item Therefore, $\bSigma^{-1}=\mathbb{E}[\bx_t\bx_t']^{- 1}=\bG^{-1}(\bI_d-\bTheta)$ if $\mathbb{E}[\bx_t\bx_t']$ is invertible.
\item One also has that $\mathbb{E}[\bv_t\bv_t']=(\bI_d-\bTheta)\bG$.
\end{enumerate}
\end{lemma}

Lemma \ref{lemma1} suggests that Proposition \ref{th:recast} is valid for any $N$ and $K<T$. 

\begin{proof}[Proof of Proposition \ref{th:recast}] 
The result follow from a direct application of Lemma \ref{lemma1} to  $\bx_{t} = (r_{1,t},1,r_{2,j,t}-r_{1,t},\br_{1,-1,t}'-\bi_{K-1}' r_{1,t})'$. See Section \ref{illustration} in the paper for an illustration.
\end{proof}

\begin{proof}[Proof of Theorem \ref{th:batch}]
Since for $b=1,\ldots,B$, $ \hat m_j(\hat\btheta_j )_{b}$ converges to $ \hat m_j( \btheta_j )_b$ at the rate $\sqrt{T} \gg \sqrt{T_b}$, standard results on batch-mean for mean estimate of a stationary sequence apply. That is, $\hat v_{j}(\hat\btheta_j )_{\!B} - \mathbb{V}[m_j(\btheta_j )] =  o_p(1)$ while $m_j(\hat\btheta_j )_{b}= m_j(\btheta_j)_{b}+o_p(1)$ and $m_j(\hat\btheta_j )_{b'}= m_j(\btheta_j)_{b'}+o_p(1)$ have zero covariance whenever $b\neq b'$ and $T\to \infty$. Furthermore, $\hat m_j(\hat\btheta_j )_{1},\ldots,\hat m_j(\hat\btheta_j )_{B-1}$ and $\hat m_j(\hat\btheta )_{B}$ behave asymptotically as i.i.d. normal random variables with mean $m_j( \btheta )$ and variance $\mathbb{V}[m_j(\btheta_j )]$ \citep{carlstein1986use}. Therefore, we can apply a Student-t test to test the mean of the sample $\{m_j(\hat\btheta_j )_{b}\}_{b=1}^B$ as in \eqref{robttest}. We refer to \citet{carlstein1986use,pedersen2020robust} and subsequent references for more details about the theory of the batch-mean method for general statistics under mixing conditions. See also \citet[Theorem 5.1 and 5.2.]{zhang2017gaussian} for  the rate of convergence of $(v_{1}(\hat\btheta_j )_{\!B},\ldots,v_{d}(\hat\btheta_j )_{\!B})'$ under the physical dependence framework.
\end{proof}

\section{Additional Results}

Table~\ref{tab:addres} reports the sample size $T$, the number benchmark assets $K$, and the number of test assets $N$ used
in the two empirical applications of Section \ref{sec:empirics} of the paper.

\begin{table}[H]
\caption{\textbf{Dimensions of the Universes in the Empirical Applications}\\
The table reports the number of observations $T$, the number of benchmark assets $K$, and test assets $N$ in the two
empirical applications of Section \ref{sec:empirics}. For the first application, the element under $N$ for a given benchmark universe equals the sum of those under the columns $K$ of the other benchmark assets, except for 2015. This is because of dual listed stocks in the SMI and Euronext 100. In the second application, the number of factors was the same every year: 5 for FF5, 6 for FF6, 
25 for FF25, 100 for FF100, and 153 for KEW and KVW. }
\label{tab:addres} 
\centering
\singlespacing
\scalebox{0.95}{
\begin{tabular}{lccccccccc}
\toprule
&\multicolumn{7}{c}{Application 1}
&\multicolumn{2}{c}{Application 2}\\
\cmidrule(lr){2-8}\cmidrule(lr){9-10}
&&\multicolumn{2}{c}{S\&P 100}&\multicolumn{2}{c}{Euronext 100}&\multicolumn{2}{c}{SMI}&\multicolumn{2}{c}{S\&P 500}\\
\cmidrule(lr){3-4}\cmidrule(lr){5-6}\cmidrule(lr){7-8}\cmidrule(lr){9-10}
Year & $T$ & $K$ & $N$ & $K$ & $N$  & $K$ & $N$ & $T$ &$N$ \\
\midrule
2007 &  244 & 99 & 118 & 98 & 119 & 20 & 197 & 251 & 496\\
2008 &  245 & 99 & 118 & 98 & 119 & 20 & 197 & 253 & 497\\
2009 &  245 & 100 & 115 & 96 & 119 & 19 & 196 & 252 & 498\\
2010 &  248 & 100 & 117 & 98 & 119 & 19 & 198 & 252 & 498\\
2011 &  248 & 100 & 115 & 95 & 120 & 20 & 195 & 252 & 496\\
2012 &  243 & 99 & 116 & 96 & 119 & 20 & 195 & 250 & 496\\
2013 &  243 & 100 & 119 & 99 & 120 & 20 & 199 & 252 & 497\\
2014 &  243 & 100 & 117 & 97 & 120 & 20 & 197 & 252 & 500\\
2015$^*$ &  246 & 101 & 118 & 99 & 120 & 20 & 199 & 252 & 495\\
2016 &  248 & 102& 118 & 98 & 122 & 20 & 200 & 252 & 502\\
2017 &  245 & 101 & 119 & 99 & 121 & 20 & 200 & 251 & 502\\
2018 &  242 & 102 & 118 & 98 & 122 & 20 & 200 & 251 & 505\\
2019 &  243 & 100 & 116 & 97 & 119 & 19 & 197 & 252& 500\\
2020 &  246 & 100 & 110 & 90 & 120 & 20 & 190 & 253 & 501\\
2021 &  248 & 101 & 103 & 83 & 121 & 20 & 184 & 252 & 504\\
2022 &  247 & 101 & 101 & 81 & 121 & 20 & 182 & 251 & 501\\																		
\bottomrule			
\end{tabular}}
\end{table}

\end{document}